\documentclass[aps,twocolumn,superscriptaddress,prl,longbibliography]
{revtex4-2}

\usepackage{xcolor}
\usepackage{amsfonts}
\usepackage{amsmath}
\usepackage{dcolumn}
\usepackage{upgreek}
\usepackage{bm}
\usepackage{tikz}
\usepackage[normalem]{ulem}
\usepackage[colorlinks=true,citecolor=blue,urlcolor=blue]{hyperref}
\usepackage{amsmath,amsfonts,amssymb,times}
\usepackage{natbib} 
\usepackage{jabbrv}
\usepackage{url}
\usepackage{etoolbox}
\makeatletter
\patchcmd{\thebibliography}{\sloppy}{\sloppy\emergencystretch=2em\hyphenpenalty=50\tolerance=2000\relax}{}{}
\makeatother
\usepackage[standard]{ntheorem}
\usepackage{braket}

\makeatletter
\newcommand*{\Rom}[1]{\expandafter\@slowromancap\romannumeral #1@}
\makeatother

\begin{document}

\title{Topological Anderson Random Laser}% Force line breaks with \\

\author{Hang-Zheng Shen}
\affiliation{Laboratory of Quantum Information, University of Science and Technology of China, Hefei 230026, China}
\affiliation{Anhui Province Key Laboratory of Quantum Network, University of Science and Technology of China, Hefei 230026, China}
\affiliation{Synergetic Innovation Center of Quantum Information and Quantum Physics, 
University of Science and Technology of China, Hefei, China}
\author{Xian-Hao Wei}
\affiliation{Laboratory of Quantum Information, University of Science and Technology of China, Hefei 230026, China}
\affiliation{Anhui Province Key Laboratory of Quantum Network, University of Science and Technology of China, Hefei 230026, China}
\affiliation{Synergetic Innovation Center of Quantum Information and Quantum Physics, 
University of Science and Technology of China, Hefei, China}
\author{Xi-Wang Luo}\email{luoxw@ustc.edu.cn}
\affiliation{Laboratory of Quantum Information, University of Science and Technology of China, Hefei 230026, China}
\affiliation{Anhui Province Key Laboratory of Quantum Network, University of Science and Technology of China, Hefei 230026, China}
\affiliation{Synergetic Innovation Center of Quantum Information and Quantum Physics, 
University of Science and Technology of China, Hefei, China}
\affiliation{Hefei National Laboratory, University of Science and Technology of China, Hefei 230088, China}
\affiliation{Anhui Center for Fundamental Sciences in Theoretical Physics,
University of Science and Technology of China, Hefei 230026, China}
\author{Zheng-Wei Zhou}\email{zwzhou@ustc.edu.cn}
\affiliation{Laboratory of Quantum Information, University of Science and Technology of China, Hefei 230026, China}
\affiliation{Anhui Province Key Laboratory of Quantum Network, University of Science and Technology of China, Hefei 230026, China}
\affiliation{Synergetic Innovation Center of Quantum Information and Quantum Physics, 
University of Science and Technology of China, Hefei, China}
\affiliation{Hefei National Laboratory, University of Science and Technology of China, Hefei 230088, China}
\affiliation{Anhui Center for Fundamental Sciences in Theoretical Physics,
University of Science and Technology of China, Hefei 230026, China}

\begin{abstract}
Topological lasers and random lasers embody two contrasting strategies for disorder management in photonics: the former suppresses disorder via protected edge transport, while the latter exploits multiple scattering for feedback. Here, we theoretically demonstrate that these seemingly incompatible paradigms can be unified through a topological Anderson random laser (TARL), where disorder itself induces a topological phase that enables robust lasing. Starting from a trivial photonic lattice, we show that engineered disorder drives the system into a topological Anderson insulator regime, generating emergent chiral edge states that serve as boundary-selective lasing channels. Remarkably, the TARL exhibits rapid mode selection toward a single edge state, producing an ultranarrow emission spectrum and enhanced slope efficiency optimized near disorder strength with maximal topological mobility gap. 
Furthermore, they exhibit single-mode-like coherence properties, deviating from Kardar-Parisi-Zhang behavior in conventional chiral topological lasers, 
while remaining significantly more robust against local perturbations than conventional random lasers.
%and maintain robustness against local perturbations and temporal noise. 
Our findings establish a disorder-enabled flexible route to topologically protected single-mode lasing and introduce a fundamentally new design principle for robust, high-coherence photonic light sources.
\end{abstract}
\date{\today}% It is always \today, today,
             %  but any date may be explicitly specified

\maketitle

%\section{\label{sec:1}Introduction}
\noindent{\bf \large Introduction}\\
Lasers are cornerstone light sources in modern science and technology, enabling applications ranging from precision metrology and imaging \cite{abbott2016observation,hell2007far, xu2021non} to optical communication and information processing \cite{marin2017microresonator, zhong2020quantum, carolan2015universal}. A central challenge in laser design is to achieve robust and stable single-mode operation with high efficiency, even in the presence of inevitable fabrication imperfections and environmental noise~\cite{Thyagarajan2011, hentschel2001attosecond, diddams2000direct}. It has been recognized that concepts from topological phases of matter have been successfully introduced into laser physics, leading to the realization of topological laser (TL)~\cite{st2017lasing, bahari2017nonreciprocal, harari2018topological}. In these devices, lasing is enforced in topologically protected edge modes, which are immune to back-scattering disorders and defects~\cite{bandres2018topological}. As a result, TLs can exhibit disorder-robust operation and enhanced slope efficiency beyond conventional cavity-array systems~\cite{ozawa2019topological, liTopologicalOnchipLasers2023}. However, TLs still face several limitations: 1) the laser spectrum may cover the entire band gap, as all edge modes exhibit comparable gain~\cite{bandres2018topological,zengElectrically2020,yangTopologicalcavity2022}; 2) due to the ideal linear dispersion of edge modes, their coherence properties fall within the Kardar-Parisi-Zhang universality class~\cite{amelioTheoryCoherenceTopological2020, PhysRevResearch.4.043207}, leading to faster decay compared to single-mode lasers. %.~\cite{Harder2024Electrically,Liu2024HighPower,leefmansTopologicalTemporallyModelocked2024,doi:10.1126/science.aba8996}.

On the other hand, random lasers~\cite{Review2025LiquidCrystal,Zhang2024Perovskite,Prokopeva2024Modeling,Dipold2024Infrared}, another class of novel lasers with contrasting characteristics, have attracted considerable attention. Unlike TLs, which are inherently immune to disorder, random lasers intentionally exploit it, and can be realized across a wide range of platforms and gain media. In such systems, multiple scattering within a disordered gain medium provides the necessary feedback for lasing~\cite{segevAnderson2013, leeTamingRandomLasers2019, adlTailoringSingleModeRandom2024}. This mechanism greatly relaxes fabrication constraints, enabling low-cost implementation and offering broad flexibility in device geometry and material choice. As such, random lasers represent a complementary paradigm to TLs: rather than suppressing disorder, they harness it to enable lasing. Nevertheless, random lasers also suffer from several drawbacks: 1) they typically exhibit multimode emission or excessively broad spectra; 
%governed by complex and often poorly understood theoretical frameworks; 
2) although careful design can lead to single-mode operation, their emission properties remain highly sensitive to perturbations~\cite{leeTamingRandomLasers2019,Shawki17}. These limitations severely restrict the practical application scenarios of random lasers~\cite{rlbook,rlreviewnew}.

TLs and random lasers thus represent two fundamentally distinct design philosophies, each with complementary strengths and limitations. These two approaches appear inherently incompatible: one seeks to eliminate the influence of disorder, while the other relies on it. A natural and important question then arises: How can these two paradigms be integrated to combine their advantages and mitigate their respective drawbacks?

Here, we demonstrate such a unification by theoretically proposing a new approach to topologically protected random lasing—the topological Anderson random laser (TARL).
%we show that such a unification is indeed possible by theoretically proposing a new approach to topologically protected random lasing, namely the topological Anderson random laser (TARL). 
This is achieved by suitably engineering gain medium in a system that supports a disorder-induced topological Anderson insulating phase\cite{liTopologicalAndersonInsulator2009, grothTheoryTopologicalAnderson2009, Shen2017, xingTopologicalAndersonInsulator2011, jiangNumericalStudyTopological2009, bernevigQuantumSpinHall2006}. While the concept of topological Anderson insulators (TAIs) has been proposed and experimentally realized in various optical platforms~\cite{stutzerPhotonicTopologicalAnderson2018, liuTopologicalAndersonInsulator2020, chenRealizationTimeReversalInvariant2024, assuncaoPhaseTransitionsScale2024, PhysRevA.106.L051301}, it remains unclear whether such phases can be exploited in the context of lasing, and if so, what their emission characteristics might be. We introduce and systematically investigate the TARL. Starting from a trivial photonic insulator, we demonstrate that disorder can drive a photonic system into a topological Anderson insulator phase, in which emerging chiral edge modes serve as the lasing channels under boundary gain. We show that, compared to conventional chiral TLs, TARLs exhibit rapid relaxation into a single edge mode, extremely narrow lasing spectra, and enhanced slope efficiency optimized {near the disorder strength with maximal topological mobility gap}; while compared to conventional random lasers, TARLs remain significantly more robust against local
perturbations. We further analyze their performance under realistic temporal noises, revealing single-mode-like coherence properties rather than Kardar-Parisi-Zhang-type behavior. Together, our results establish TARLs as a robust and experimentally promising platform that merges the advantages of both topological and random lasers. \\

\begin{figure}[t]

\includegraphics[width=\linewidth]{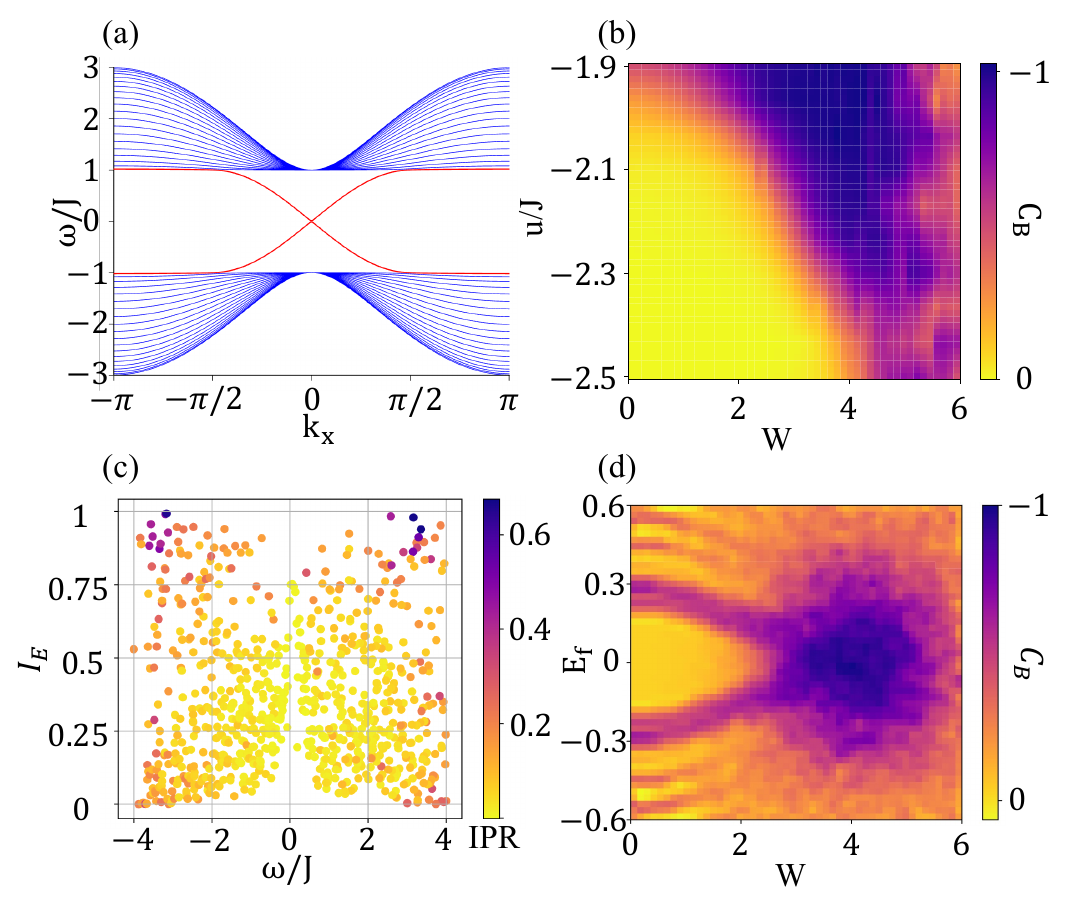}% Here is how to import EPS art
\caption{\label{fig:1} Band structure and topological phase characteristics.(a) Band structure of a cylindrical QWZ lattice with width $N_{y}=20$ and parameter $u=-1$. 
(b) Phase diagram in the $w$-$u$ plane determined by the Bott index $C_B$ with $E_{f}=0$, the TAI phase appears in the purple area.
(c) Eigenvalue distribution of a finite TAI system with $N_{x}=20, N_{y}=20, u=-2.1$, and disorder strength $W=3.5$. The horizontal axis shows the eigenvalue $\omega$, and the vertical axis indicates the edge population $I_E$. The color map represents the IPR, characterizing the localization of each eigenstate.
(d) Bott index $C_B$ evaluated over different disorder strength $W$ and Fermi energy $E_{f}$, showing the closure of the trivial band gap (orange) followed by the emergence of a disorder-induced topological mobility gap (purple).
%In panels (b) and (d), the data are processed using filtering and interpolation.
}

\end{figure}

%In this work, we introduce and systematically investigate a topological Anderson random laser (TARL). Using a topological Anderson insulator model based on the Qi-Wu-Zhang model~\cite{Qi2006}, we demonstrate that disorder can drive a photonic system into a topological Anderson insulator phase whose chiral-like edge modes serve as the lasing channels under boundary gain. We show that, compared with conventional chiral TLs, TARLs exhibit rapid relaxation into a single edge mode, extremely narrow realization-averaged spectra, and enhanced slope efficiency optimized near the maximal topological Anderson gap. We further analyze their robustness against local perturbations and their performance under realistic white-noise conditions, revealing single-mode-like coherence properties rather than the Kardar-Parisi-Zhang-type behavior~\cite{amelioTheoryCoherenceTopological2020, PhysRevResearch.4.043207} typical of multimode TLs. Overall, our results establish TARLs as a robust and experimentally promising platform that combines the advantages of TLs and random lasers.
%\section{Results}
%\subsection{\label{sec:2}THE MODEL}
\noindent{\bf \large Results}\\
\noindent{\bf \small The model}\\ 
To explore the properties of our proposed TARL in a simple way while preserving its essential features, we focus on the disorder-driven topological Anderson phase from the trivial phase of the Qi-Wu-Zhang (QWZ) model~\cite{Qi2006,Asbóth2016}, and our results can be applied to other models straightforwardly. 
%The topological Anderson insulator (TAI) phenomenon has been extensively studied in various contexts~\cite{liTopologicalAndersonInsulator2009, grothTheoryTopologicalAnderson2009, Shen2017, xingTopologicalAndersonInsulator2011, jiangNumericalStudyTopological2009, bernevigQuantumSpinHall2006}, and has been realized in photonic systems~\cite{stutzerPhotonicTopologicalAnderson2018, liuTopologicalAndersonInsulator2020, chenRealizationTimeReversalInvariant2024, cuiPhotonic2Topological2022, renRealizationGappedUngapped2024, PhysRevA.105.043514, daiProgrammableTopologicalPhotonic2024}. Since most experimental realization of the photonic topological Anderson insulator are based on Haldane-like model~\cite{PhysRevLett.61.2015}, we also show a more practical model based on Haldane model in Sec.~1 of the Supplemental Document. 
%The TL constructed from the QWZ model is taken as a representative of conventional chiral TLs. 
The Hamiltonian of the QWZ model with disorders is given by:
\begin{eqnarray}
\label{eq:1}
        H &=& J\sum_{m,n} \left(\psi^\dagger_{m+1,n} \frac{\sigma_z+i\sigma_x}{2}\psi_{m,n}+h.c.\right)\nonumber\\
         &&+J\sum_{m,n} \left(\psi^\dagger_{m,n+1} \frac{\sigma_z+i\sigma_y}{2}\psi_{m,n}+h.c.\right)\nonumber\\
        &&+u\sum_{m,n}  \psi^\dagger_{m,n} \sigma_z\psi_{m,n}+\sum_{m,n,s}  \Delta_{m,n,s}\psi^\dagger_{m,n,s}\psi_{m,n,s} 
\end{eqnarray}
where $\psi^\dagger_{m, n}=(\psi^\dagger_{m, n,A},\psi^\dagger_{m, n,B})$ denote the creation operator at site $(x=m, y=n)$. $s=A,B$ represent the sublattice states and $\sigma_i$ are the Pauli matrices. $J$ represents the hopping amplitude and we set $J=1$ as the energy unit. $u$ denotes the staggered on-site potential and $\Delta_{m,n,s}$ is the on-site disorder. In the clean limit $\Delta_{m,n,s}=0$, the topological properties are determined by the ratio $u/J$. When $|u| > 2$, the system is trivial with Chern number $C= 0$; when $|u|< 2$ (excluding $u = 0$), the system is nontrivial with Chern number $C = \text{sgn}(u)$, whose bulk band and edge states in a cylindrical geometry is shown in Fig.\ref{fig:1}(a).

%The TAI phase emerges under the condition $|u| > 2$ through the introduction of disorder. 
In our model, disorder is introduced and quantified by adding a site and sublattice independent random potential uniformly distributed in the interval $\Delta_{m,n,s}\in[-W/2, W/2]$. The parameter $W$ represents the disorder strength of the system. 
The topological Anderson model lacks a well-defined momentum-space formulation. The topological Chern number is replaced by the Bott index $C_B$, which is the disorder averaged local Chern marker at the center of the system for the occupied states below the Fermi energy $E_f$ (see SI for details). Here, we consider the disordered system with open boundary conditions and evaluate the Bott index $C_B$.  
%The validity of this approach is discussed in Sec.~2 of the Supplemental Document. It is defined as follows:
%\begin{equation}
%    \label{eq:2}
%    C(m,n) = -2\pi i\sum_s\bra{m,n,s}[PxP,PyP] \ket{m,n,s}
%\end{equation}
%where, $P$ is the projection operator that projects states onto the occupied subspace below an effective Fermi energy $E_f$.
The associated phase diagram in the $u$-$W$ plane with $E_f=0$ is shown in Fig.~\ref{fig:1}(b).
It can be seen that, starting from the trivial phase with 
$|u|>2$, the TAI phase (with $|C_B|=1$) emerges as we increase the disorder to medium strength, and further increasing the disorder strength drives the system to the trivial Anderson localization regime.
It is worthy to note that when $|u|$ is too large, the system transitions directly into Anderson localization without entering the topological Anderson phase as disorder increases.

\begin{figure*}[t]
\includegraphics[width=\linewidth]{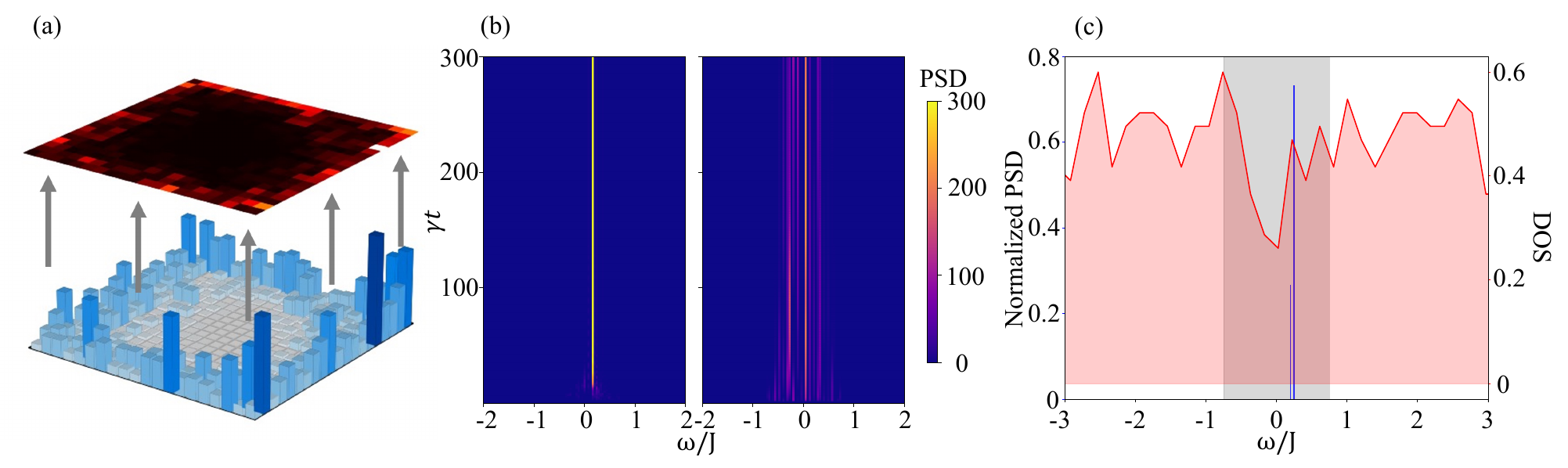}
\caption{\label{fig:2} Single-mode lasing behavior in TARLs. (a) Schematic demonstration of how the TARL supports a single edge mode that dominates the lasing dynamics.
(b) Time-resolved power spectral density (PSD) for the TARL (left panel) and a conventional QWZ-based TL (right panel), highlighting the rapid single-mode selection in the TARL case.
(c) The normalized PSD (blue bars) averaged over initial fluctuations and the density of states (red line) of the TARL system. The gray shaded region marks the pumping frequency window. The parameters for the TARL (conventional TL) are: $u=-2.1$ and $W=3.5$ ($u=-1$ and $W=0$). Other parameters are {$P=0.5$, $\gamma=0.1$}.}
\end{figure*}

In the following discussion, we will focus on the case with $u = -2.1$. To show how the eigenstates are distributed for the TAI phase,
in Fig.\ref{fig:1}(c), we provide the scatter plot of eigenenergy $\omega$ versus edge population $I_E=\sum_{m,n\in \text{edge}}\langle \psi^\dagger_{m,n}\psi_{m,n}\rangle$ (the summation includes two layers nearest to the boundary) as well as the inverse participation ratio (IPR)~\cite{IPR1,IPR2} of all the eigenstates at disorder strength $W = 3.5$. Only a few states near $\omega\approx 0$ exhibit both high edge population and low IPR, identifying them as extended chiral edge modes; the spectrum is otherwise dominated by strongly localized states and bulk states. Moreover, by varying $E_f$, one can observe corresponding changes in the Bott index, providing a clear indication of the presence and location of mobility gaps~\cite{liTopologicalAndersonInsulator2009}. As shown in Fig.\ref{fig:1}(d), increasing disorder gradually closes the trivial band gap and opens a topological gap. When the disorder goes to the strong limit, all states become localized, the mobility band gap closes again, and the system enters the Anderson localization regime.

Based on the semiclassical theory of lasers~\cite{Thyagarajan2011,harari2018topological}, we introduce an effective mean-field optical amplitude $\Psi_{m,n}=\langle \psi_{m,n}\rangle$ to describe the optical field. The gain medium is modeled by adding a non-Hermitian gain term to the Schr\"{o}dinger-like equation governing the field evolution
%, while intrinsic losses are introduced through a dissipation term. Drawing on previous studies of topological and random lasers, we apply the gain uniformly along the entire boundary, targeting modes near the energy gap.
\begin{equation}
    \label{eq:3}
    i\partial_t \Psi_{m,n} = (H\Psi)_{m,n} + i(\frac{P_{m,n}(\omega)}{1+\beta \langle\Psi|\Psi\rangle}-\gamma)\Psi_{m,n}
\end{equation}
where $P_{m,n}(\omega)$ determines the spatial and spectral characteristics of the gain~\cite{Secli:19}. To select the topological Anderson chiral modes, we set $P_{m,n}(\omega)=P$ if $(m,n)$ belongs to the boundary sites (the outmost layer) and $\omega\in [-0.75J, 0.75J]$ (i.e., the spectral window of gain is approximately twice the energy gap), $P_{m,n}(\omega)=0$ otherwise. 
%to be uniformly distributed along the boundary and 
%$\omega\in [-0.75J, 0.75J]$, 
%targeting the topological Anderson chiral modes. 
The parameter $\gamma$ represents intrinsic loss, while $\beta$ characterizes the level of gain saturation. We numerically evolve the system from a weak random initial field across the lattice.
Without loss of generality, all discussions below are restricted to open-boundary systems with $20 \times 20$  unit cells. \\

%\subsection{\label{sec:3}Evaluation of lasing properties}
\noindent{\bf \small Evaluation of lasing properties} \\
We begin our evaluation with the behavior of the system in the absence of temporal noise. In this case, after competition among the edge modes, the system randomly locks into one of them and subsequently maintains stable single-mode operation, as shown in Fig. \ref{fig:2}(a). 
%\subsubsection{\label{sec:2A}Mode evolution and distribution}
The mode dynamics are characterized using power spectral density (PSD) distributions as a function of time~\cite{ozawa2019topological,Thyagarajan2011}. The PSD in our system can be obtained by calculating $|\braket{v^{(j)}|\Psi}|^2$, where $|v^{(j)}\rangle$ is the eigenmode of the Hamiltonian with eigenfrequency $\omega=\omega_j$. As shown in Fig.~\ref{fig:2}(b), the TARL rapidly reaches to a stable single-mode steady state. As comparison, the conventional TL 
%(with $u=-1,W=0$) 
still undergoes mode competition even at
$\gamma t=300$, this ultra-slow relaxation behavior~\cite{secliTheoryChiralEdge2019} hinders practical application which is circumvented in our system. 
To simplify the numerical simulation in Fig.~\ref{fig:2}, we set $\beta=1$ to obtain fast saturation.
%Moreover, 
The TARL exhibits lasing confined to at most two adjacent modes under repeated realizations (with different random initial field), resulting in an extremely narrow realization-averaged normalized PSD, where the normalization is performed both for the single-realization PSD and for the number of realizations, as shown in Fig.~\ref{fig:2}(c).
Compared to the conventional chiral TL, nearly all edge modes exhibit a substantial probability of becoming the final lasing mode, leading to the realization-averaged PSD covering the entire topological gap~\cite{secliTheoryChiralEdge2019}. 
In the presence of temporal noise (whose effects will be discussed latter), a narrower PSD would result in narrower laser spectral broadening~\cite{bandres2018topological,zengElectrically2020,yangTopologicalcavity2022}.

\begin{figure}[t]
\includegraphics[width=\linewidth]{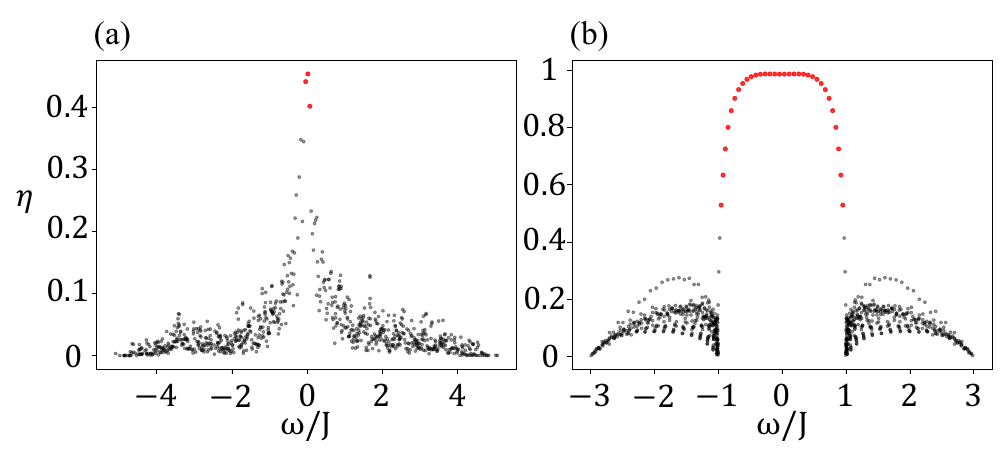}
\caption{Eigenfrequencies and normalized GMO factors $\eta_j$ of all modes for (a) the TARL with $u=-2.1$, $W=3.5$ and (b) the conventional TL with $u=-1$, $W=0$. Black (red) circles denote bulk (edge) modes. \label{fig:n1}}
\end{figure}

To provide a theoretical explanation and quantify the likelihood that a given state is ultimately selected, we evaluate introduce the normalized gain–mode overlap (GMO) factor as 
\begin{eqnarray}
    \eta_j =
\frac{(\int \mathrm{d}\mathbf{r}\, G(\mathbf{r})\,|v^{(j)}(\mathbf{r})|)^2}
{\int \mathrm{d}\mathbf{r} |G(\mathbf{r})|^2}
\end{eqnarray}
where $G(\mathbf{r})$ denotes the spatial profile of the boundary gain and $v^{(j)}(\mathbf{r})$ is the corresponding profile of the $j$-the mode, $\mathbf{r}=(m,n,s)$. The modulus is taken to remove any phase dependence. This factor is bounded by unity. Its relative magnitude across different modes characterizes their competitiveness during the evolution, with larger values indicating a higher likelihood of becoming the dominant lasing mode.
%For our model, the factor is:
%$$
%\eta[v] = \frac{1}{2N_{g}}\left(\sum_{m,n\in \text{edge}} |v_{m,n}|\right)^2,
%$$
%where $N_g$ denotes the number of sites, and the factor of 2 originates from the two sublattices.
As shown in Fig.~\ref{fig:n1}(a), disorder strongly disrupts the spatial similarity among edge modes in TARLs. As a result, only a small number of modes remain spatially extended along the boundary and retain substantial overlap with the gain region. Moreover, these modes exhibit separated normalized GMO factors. The reduced number of effectively competing modes leads to a shorter relaxation time and a significantly narrower ensemble-averaged PSD.
While for conventional TLs (see Fig.~\ref{fig:n1}(b)), a large number of edge modes possess similar spatial profiles. Consequently, they exhibit comparable normalized GMO factors and therefore similar amplification strengths. This results in prolonged and complex mode competition, which gives rise to a broad PSD.

%\subsubsection{\label{sec:2B}Laser intensity behavior}

\begin{figure}[t]
\includegraphics[width=\linewidth]{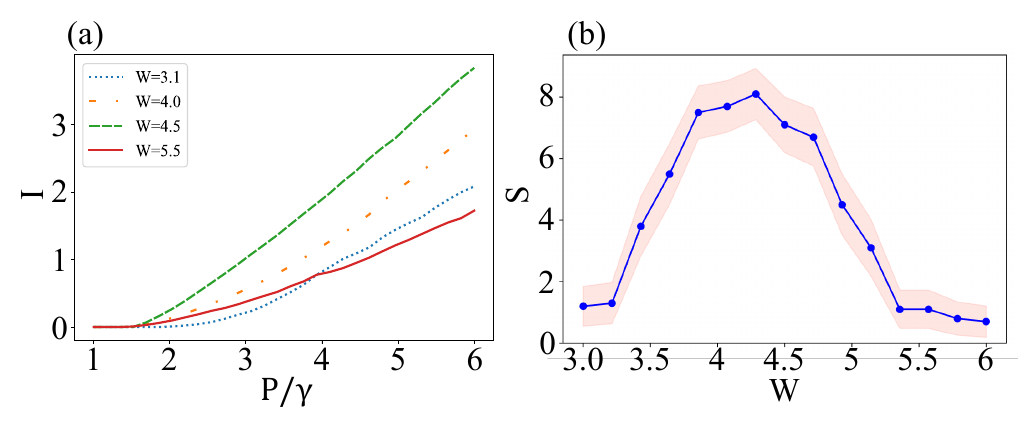}
\caption{\label{fig:3} Slope efficiency of the TARL. (a) Laser output intensity $I$ as a function of pump strength $P/\gamma$ for different disorder strengths $W$. (b) Slope efficiency $S$ as a function of disorder strength $W$. The results are averaged over 100 realizations of disorder reconfiguration, with the shaded region indicating the 95\% confidence interval. The efficiency is maximized near the disorder value $W\simeq 4$ that corresponds to the largest topological mobility gap [see Fig.~\ref{fig:1}(c)]. Other parameters are the same as that in Fig.~\ref{fig:2}.
}
\end{figure}

Now we consider the output efficiency of the system. The laser intensity is defined as $I = \braket{\Psi|\Psi}/N_{g}$, where the summation runs over the entire system and $N_{g}$ is refers to the number of sites where gain is applied. We examine the relationship between $I$ and the gain coefficient, which is characterized by the normalized parameter $P/\gamma$. For TARLs under various disorder strengths, we vary the gain coefficient $P$ and compute the ensemble average of the output intensity $I$ after the system evolves to a steady state. As shown in Fig.~\ref{fig:3}(a), we find that the threshold gain is relatively consistent across different disorder strengths and slightly exceeds $\gamma$. Moreover, the relationship between $I$ and $P$ is approximately linear. Therefore, we use the slope efficiency to characterize the output efficiency of the system $ S = \frac{\Delta I}{\Delta P}$.
For TARLs, we compute the disorder-configurations-averaged slope efficiency under various disorder strengths, as shown in Fig.~\ref{fig:3}(b). As expected, we observe a non-monotonic behavior: as the disorder strength increases, the slope efficiency initially improves and then declines.
The disorder strength at which the laser output intensity reaches its maximum coincides with the disorder strength where the topological mobility gap is maximized in Fig.\ref{fig:1}(c). In the topological phase, the presence of topologically protected edge states leads to a higher slope efficiency compared to the trivial phase. When the disorder strength lies within this optimal range, the system is furthest from the trivial phase, and the topological protection of the edge states is strongest, resulting in the highest slope efficiency.
In conventional TLs, increasing disorder typically usually leads to a monotonic decrease in slope efficiency~\cite{st2017lasing, bahari2017nonreciprocal, harari2018topological}.

\begin{figure}[t]
\includegraphics[width=\linewidth]{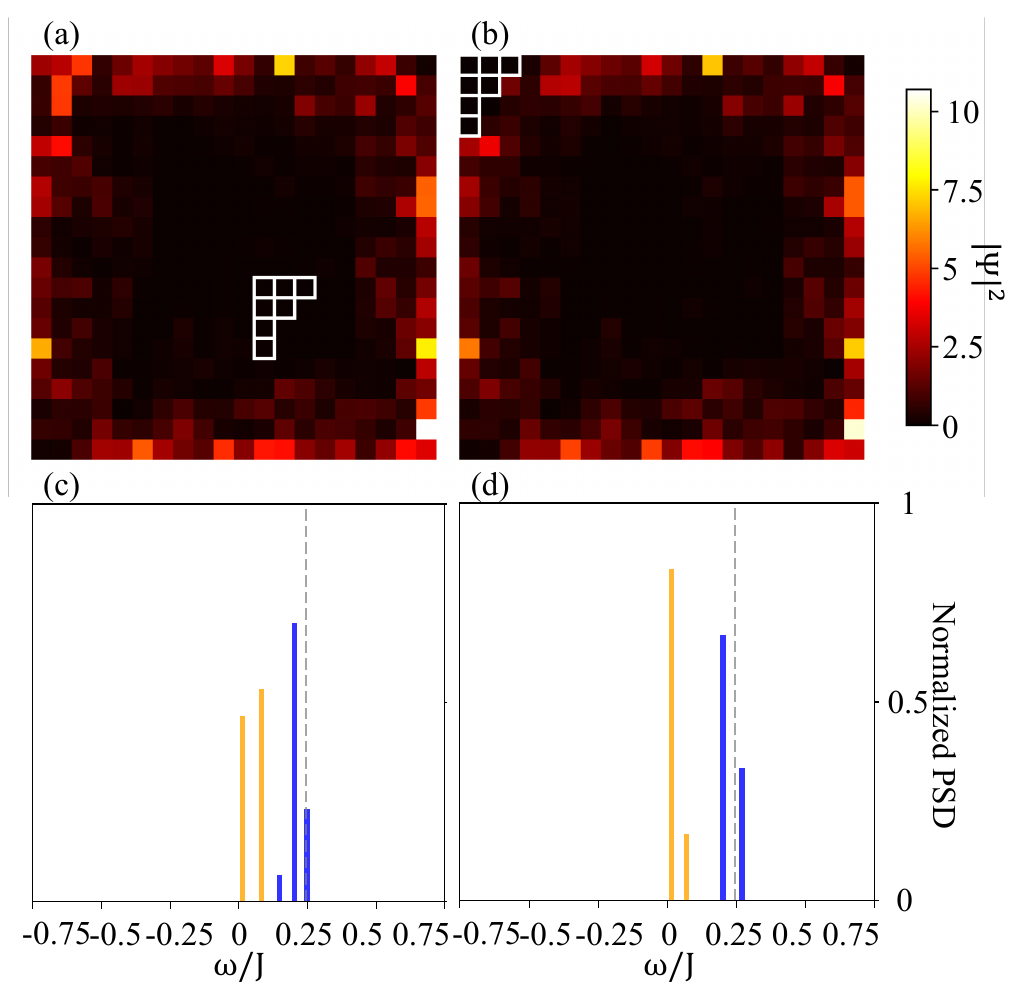}
\caption{\label{fig:4} Robustness of the TARL against local perturbations.  (a) and (b) the real-space patterns of the steady laser with defects (white squares) introduced in the bulk and boundary, respectively. (c) The normalized PSDs (blue bars and yellow bars) corresponding to the defects in (a) and (b), respectively. (d) The normalized PSD disorder reconfigurations of $3$ sites (blue bars) and $20$ sites (yellow bars) compared with the disorder configuration in Fig.~\ref{fig:2}. The gray dashed line marks the position of the PSD peak in Fig.~\ref{fig:2}(c). Other parameters are the same as in Fig.~\ref{fig:2}.
}
\end{figure}

%\subsubsection{\label{sec:2C}Robustness against local perturbations}
To investigate the robustness of the TARL, we evaluate both the single-shot emission and the changes of the initial-fluctuation averaged PSD in the presence of
two types of imperfections: fabrication defects and disorder reconfiguration.
We find that the single-mode property in each single-shot emission is guaranteed by topological protection, while the emission spectra may be slightly modified by these imperfections.
We first consider the impact of defects by introducing them either in the bulk or along the boundary of the system. The corresponding single-shot emission profiles are shown in Figs.~\ref{fig:4}(a),(b), and the initial-fluctuation-averaged PSD are shown in Fig.~\ref{fig:4}(c). 
We see that the system still exhibits single-mode lasing enabled by topological edge states, and the averaged PSD only changes slightly near $\omega=0$, indicating that the edge-mode lasing is free from the detailed shape of the boundary. Note that boundary defects lead to more pronounced spectral shifts than bulk defects due to the larger overlap with lasing edge mode. 
We now consider the effects of disorder reconfiguration. Similarly, we find the single-mode lasing is always maintained. 
%The corresponding changes in the normalized PSD are again examined. The normalized PSD does not change dramatically under disorder reconstruction. We next present the PSDs of two representative reconstructed systems, in which 3 and 20 sites are randomly reconstructed, respectively. 
Meanwhile, the PSD is hardly changed for reconfiguring a small number of sites, and only slightly shifts for reconfiguring a large number of sites [see Fig.~\ref{fig:4}(d) for the results with 3-site and 20-site disorder reconfiguration]
%with 3-site disorder reconfiguration and Fig.~\ref{fig:4}(b) with 20-site disorder reconfiguration]. 
In summary, the interplay between topology and disorder renders the TARL robust against local perturbations (see SI for more details). \\

%\subsection{\label{sec:4}Effects of temporal noises}
\noindent{\bf \small Effects of temporal noises} \\
The analysis so far has focused on the behavior of TARLs without involving realistic temporal noise, which can significantly affect both the spectral characteristics and coherence properties of the laser emission. To comprehensively assess the performance of TARLs under practical conditions, we extend our analysis to include temporal noise in the system dynamics and examine the emission spectrum and coherence properties. 
The dynamics is governed by
\begin{eqnarray}
\label{eq:6}
i\partial_t \Psi_{m,n} &=& (H\Psi)_{m,n} 
+ i\left( \frac{P_{m,n}(\omega)}{1 + \beta \langle\Psi|\Psi\rangle} - \gamma \right)\Psi_{m,n} \nonumber\\
&& + \sqrt{2D_{m,n}}\,{\xi}_{m,n}
\end{eqnarray}
We focus on the white noise with $\langle \xi^*_{m,n}(s,t)\, \xi_{m^{\prime},n^{\prime}}(s^{\prime},t^{\prime}) \rangle = \delta_{m,m^{\prime}} \delta_{n,n^{\prime}} \delta_{s,s^{\prime}} \delta(t - t^{\prime})$ with $D_{m,n}$ the spatial distribution coefficient of the noise. Following the Wigner approach, we set $D_{m,n} = (1+\delta_{(m,n),\text{edge}})\gamma/2$ such that the noise is stronger on the boundary, reflecting the influence of spontaneous emission induced by gain. This spatial dependence captures the physical origin of noise more realistically in active regions of the laser. The saturation parameter is chosen as $\beta = 1\times 10^{-3}$~\cite{secliTheoryChiralEdge2019,amelioTheoryCoherenceTopological2020}.

\begin{figure}[t]
\includegraphics[width=\linewidth]{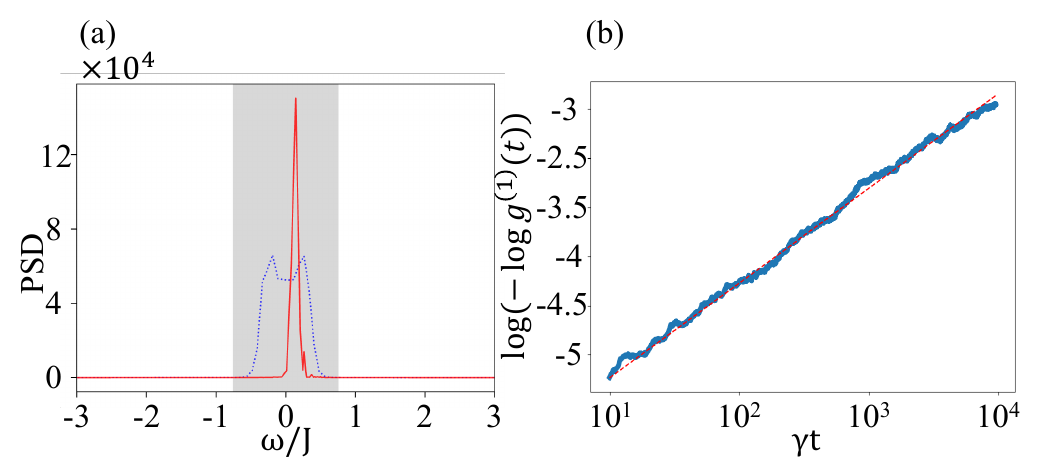}
\caption{\label{fig:5} Spectral and coherence properties under white temporal noise. (a) Comparison of the trajectory-ensemble averaged PSD for the TARL (red line) with {$W=3.5, u=-2.1$} and a conventional QWZ TL (blue line) with {$W=0, u=-1$}, demonstrating the narrower spectrum of the TARL. (b) Log-log plot of the coherence function $g^{(1)}(t)$ for the TARL. The black line is a linear fit, yielding a decay exponent $0.9656$ and indicating single-mode-like exponential decay. Other parameters: $\gamma = 0.1$, $P = 0.5$.}
\end{figure}

We perform trajectory ensemble averaging of the PSD after multiple evolutions to stable intensity, which reflects the actual emission spectrum of the system. We find that, compared to conventional chiral TLs, TARLs exhibit a significantly narrower ensemble-averaged spectrum [see Fig.~\ref{fig:5}(a)]. This is primarily because there are fewer competitive edge modes in TARLs, where no well-defined dispersion relation exists among edge states, and the spatial similarity is also significantly reduced, resulting in fewer modes that match well with the gain profile. Therefore, spectral broadening caused by multi-mode competition is greatly suppressed. While in conventional TL, due to the comparable gain and linear dispersion of the edge modes, the system cannot remain locked in a single mode in the presence of noise; switches between adjacent edge modes lead to spectral broadening which manifests as the broadening of ensemble-averaged PSD.  

To investigate the coherence properties of the emission, we can define the space-time coherence function as $g^{(1)}(\mathbf{r},t)= \frac{|\langle \Psi^*(0,0)\Psi(\mathbf{r},t)\rangle_\text{ens}|}{\sqrt{\langle \Psi^*(0)\Psi(0)\rangle_\text{ens}\langle \Psi^*(\mathbf{r},t)\Psi(\mathbf{r},t)\rangle_\text{ens}}}$~\cite{Born_Wolf_2019}.
For the TARL with a single-mode steady state, we have
$\Psi_{m,n} \simeq A(t) v_{m,n}^{(j)} \exp [-i\omega_jt-i\phi(t)]$,
where $j$ denotes the final emission eigenmode $v_{m,n}^{(j)}$ with eigenfrequency $\omega_j$, $A$ and $\phi$ represent the amplitude and phase. Therefore,
the coherence reduced to purely temporal
$g^{(1)}(t) = \frac{|\langle c^*(0)c(t)\rangle_\text{ens}|}{\sqrt{\langle c^*(0)c(0)\rangle_\text{ens}\langle c^*(t)c(t)\rangle_\text{ens}}}$
with $c(t)= A \exp [-i\phi]$ and $\langle \cdot \rangle_\text{ens}$ the ensemble averaging, we have adopted the co-moving frame by dropping the coherent phase $\omega_jt$. The results are shown in Fig.~\ref{fig:5}(b), and we find that $g^{(1)}$ takes the form~\cite{PhysRevLett.99.126403,Thyagarajan2011}.
\begin{eqnarray}
    g^{(1)}(t)= \exp[-({t}/{\tau})^{\zeta}].
\end{eqnarray}
The fitted coefficient reads $\zeta = 0.9656$, indicating that the coherence decay follows exponential behavior as in a single-mode laser. 
This is different from the conventional TLs which exhibit a crossover from exponential to faster Kardar–Parisi–Zhang type coherence decay as the system size increases~\cite{fontaineKardar2022,amelioTheoryCoherenceTopological2020}. 
%This behavior is more similar to the coherence properties of a single-mode laser rather than the KPZ-type behavior observed in TLs~\cite{fontaineKardar2022,amelioTheoryCoherenceTopological2020}.

The coherence behavior can be approximately derived from the dynamical equation by 
assuming a single-mode ansatz $\Psi_{m,n}(t) =
    A(t) v^{(j)}_{m,n} e^{-i\omega_jt-i\phi(t)} + \sum_{j'\neq j} \delta c_{j'}(t)e^{-i\omega_{j'}t}v^{(j')}_{m,n}$
%\begin{equation}
%    \label{eq:16}
%    \Psi_{m,n}(t) =
%    A(t) v^{(j)}_{m,n} e^{-i\omega_jt-i\phi(t)} + \sum_{j'\neq j} \delta c_{j'}(t)v^{(j')}_{m,n},
%\end{equation}
with $|\delta c_j|\ll |A|$.
We can obtain an effective phase evolution equation
\begin{equation}
    \label{eq:17}
    \partial_t \phi
    \simeq   \frac{\sqrt{2\gamma
    }}{A}\xi_\phi -\gamma \sum_{j'\neq j} \text{Im}\left[\langle v^{(j)}|G|v^{(j')}\rangle \frac{\delta c_{j'}e^{-i\omega_{j'}t}}{Ae^{-i\omega_j t-i\phi}}\right],
\end{equation}
where the first term on the right-hand side denotes the effective white noise while the second term represents the correction induced by mode fluctuations which is typically very weak. Therefore, we obtain an roughly exponential decay of temporal coherence (see SI for more details).\\

%\section{discussion}
\noindent{\bf \large Discussion}\\
In summary, we have demonstrated that the conceptual divide between disorder-suppressing TLs and disorder-exploiting random lasers can be resolved within a unified framework. By introducing the topological Anderson random laser (TARL), we show that disorder, when properly engineered, can simultaneously induce topological protection and provide an effective mechanism for mode selection. Starting from a trivial photonic lattice described by the QWZ model, we established that on-site disorder drives the system into a topological Anderson insulator phase, where emergent chiral edge modes become the preferred lasing channels under boundary gain.

Unlike conventional chiral TLs, in which multiple edge states compete across the band gap, the TARL rapidly relaxes into a single edge mode. This leads to strongly suppressed multimode dynamics, an ultranarrow emission spectrum, and enhanced slope efficiency. Notably, the efficiency exhibits a nonmonotonic dependence on disorder strength and is optimized near the disorder regime where the topological mobility gap is maximal, directly linking lasing performance to disorder-induced topological protection. Importantly, the TARL retains stable edge-state emission under additional local defects and partial rearrangements of the disorder landscape, highlighting its intrinsic tolerance to structural imperfections compared to conventional random lasers.

Moreover, the TARL also displays coherence behavior  fundamentally distinct to the Kardar–Parisi–Zhang universality class scaling associated with conventional TLs. Under temporal white noises, it maintains a narrow ensemble-averaged spectrum and shows nearly exponential, single-mode–like coherence decay. This difference originates from rapid collapse into a unique edge mode without a well-defined linear dispersion, which suppresses boundary phase fluctuations and nonlinear mode competition. 

From an experimental perspective, the required ingredients (i.e., disordered photonic lattices exhibiting topological Anderson phases combined with rare-earth or semiconductor gain media) are compatible with existing optical waveguide and resonator platforms~\cite{solidstate}, suggesting realistic implementation pathways. Moreover, while we have focused on the QWZ model, the results apply directly to other models (e.g., Haldane-like model, see SI for more details). Overall, our results establish a new disorder-enabled design principle for laser engineering: rather than merely mitigating imperfections, disorder can be harnessed to induce topological protection, enforce mode selectivity, and enhance coherence simultaneously. The TARL therefore provides a promising route toward robust, high-efficiency, single-mode photonic light sources and opens a broader avenue for integrating topology and complexity in active photonic systems. \\

%\section{materials and methods}
%\noindent{\bf \large Materials and Methods}\\
%The primary methods of our study were numerical simulations supplemented by analytical derivation. The characteristic properties of the TAI, including the energy band structure and the Bott index, as well as the dynamical behaviors of the TARL such as its temporal evolution and PSD, are obtained through numerical simulations. Based on the numerical results of the TARL dynamics in the presence of temporal noise, we assume the small fluctuations on top of the dominant lasing mode and further derive analytically the phase and amplitude evolution equation, which correctly predict the single-mode-like coherence properties. More details are presented in the supplementary information.\\

\noindent{\bf \large Data availability}\\
All data supporting the findings of this study are included in the article and supplementary information.\\

\noindent{\bf \large Competing interests}\\
The authors declare no competing financial interests.\\

\noindent{\bf \large Author Contributions}\\
H.S., X.L. and Z.Z. were responsible for development of the physical content and preparation of the manuscript. H.S. performed the numerical simulation. All authors contributed to the coordination and execution of this collaboration.\\

\noindent{\bf \large Acknowledgments}\\
We thank Zong-Quan Zhou for helpful discussion. This work is supported by the Natural Science Foundation of China (Grants No. 12474366, No. 12574544) and Quantum Science and Technology-National Science and Technology Major Project (Grants No. 2021ZD0301200). X.L. also acknowledges support from the USTC start-up funding.

%\clearpage
%\bibliographystyle{apsrev4-2}
%\bibliography{citation}

%apsrev4-2.bst 2019-01-14 (MD) hand-edited version of apsrev4-1.bst
%Control: key (0)
%Control: author (72) initials jnrlst
%Control: editor formatted (1) identically to author
%Control: production of article title (-1) disabled
%Control: page (0) single
%Control: year (1) truncated
%Control: production of eprint (0) enabled
%

\onecolumngrid
\setcounter{equation}{0}
\setcounter{figure}{0}
\renewcommand{\theequation}{S\arabic{equation}}
\renewcommand{\thefigure}{S\arabic{figure}}

\section{Supplementary Information}

\subsection{Calculation of the local chern marker}
We characterize the Chern number by calculating the averaged local Chern marker at the center of the sample. The local Chern marker is defined as:
\begin{equation}
\label{eq:2}
C(m,n) = -2\pi i\sum_s \bra{m,n,s}[P_\text{occ}xP_\text{occ},P_\text{occ}yP_\text{occ}] \ket{m,n,s},
\end{equation}
where $P_\text{occ}$ is the projection operator that projects states onto the occupied subspace below an effective Fermi energy $E_f$.

\begin{figure}
\centering
\includegraphics[width=0.8\linewidth]{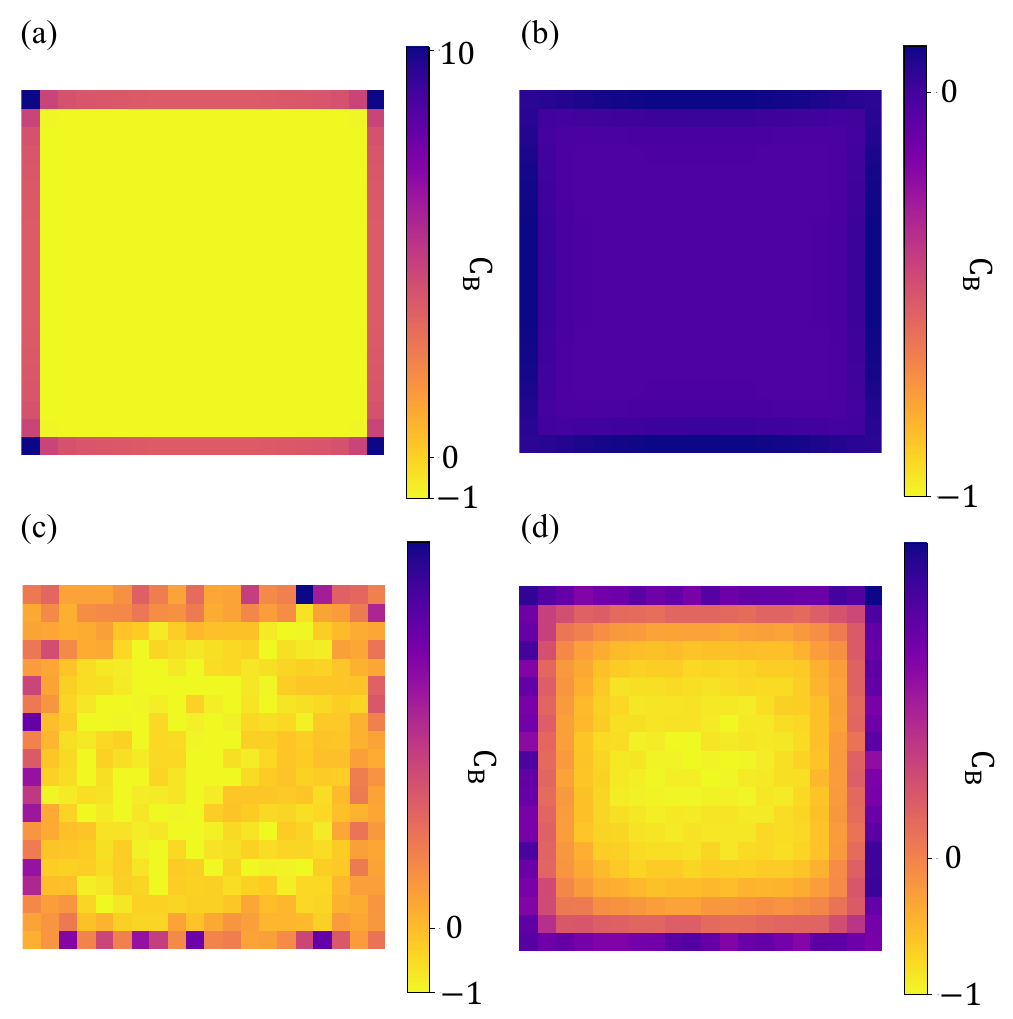}
\caption{Numerical characterization of the LCM in clean and disordered systems. (a) LCM for the QWZ model in the topological phase ($u=-1$). The central region exhibits a uniform plateau, showing excellent agreement with the quantized Chern number $C=-1$. (b) Local Chern marker for the clean QWZ model in the trivial phase ($u=-2.1$), where the values at the lattice center vanish. (c) LCM for a single disordered ($u=-2.1$, $W=3.5$) realization in the TAI phase. Significant spatial fluctuations are observed due to the presence of disorder. (d) Configuration-averaged LCM over 50 disordered realizations. The ensemble average restores spatial uniformity at the lattice center, where the converged value accurately reflects the topological nature of the TAI phase. All calculations are performed on a finite lattice of size $20 \times 20$.}
\label{figap:3} 
\end{figure}

These central values are spatially uniform and consistent with the bulk topological invariants in the thermodynamic limit (see Fig.~\ref{figap:3}(a)(b)). In the presence of disorder, i.e., in the TAI phase, the Bott index is obtained through a configurational average of the local Chern marker. Although individual disordered configurations yield an inhomogeneous local Chern marker distribution, the averaged central value converges to a well-defined plateau, which faithfully captures the topological nature of the disordered system [Fig.~\ref{figap:3}(c)(d)].

On the other hand, although the gap in the TAI differs from that in conventional topological insulators, varying $E_f$ reveals different values of the local Chern marker under the same system parameters. This behavior indicates the presence of a so-called mobility gap~\cite{liTopologicalAndersonInsulator2009S}. Such a mobility gap has been widely used in previous studies to characterize the eigenvalue distribution associated with edge states~\cite{stutzerPhotonicTopologicalAnderson2018S, liuTopologicalAndersonInsulator2020S, chenRealizationTimeReversalInvariant2024S, assuncaoPhaseTransitionsScale2024S, PhysRevA.106.L051301S}.

\begin{figure}[t]
\centering
\includegraphics[width=0.8\linewidth]{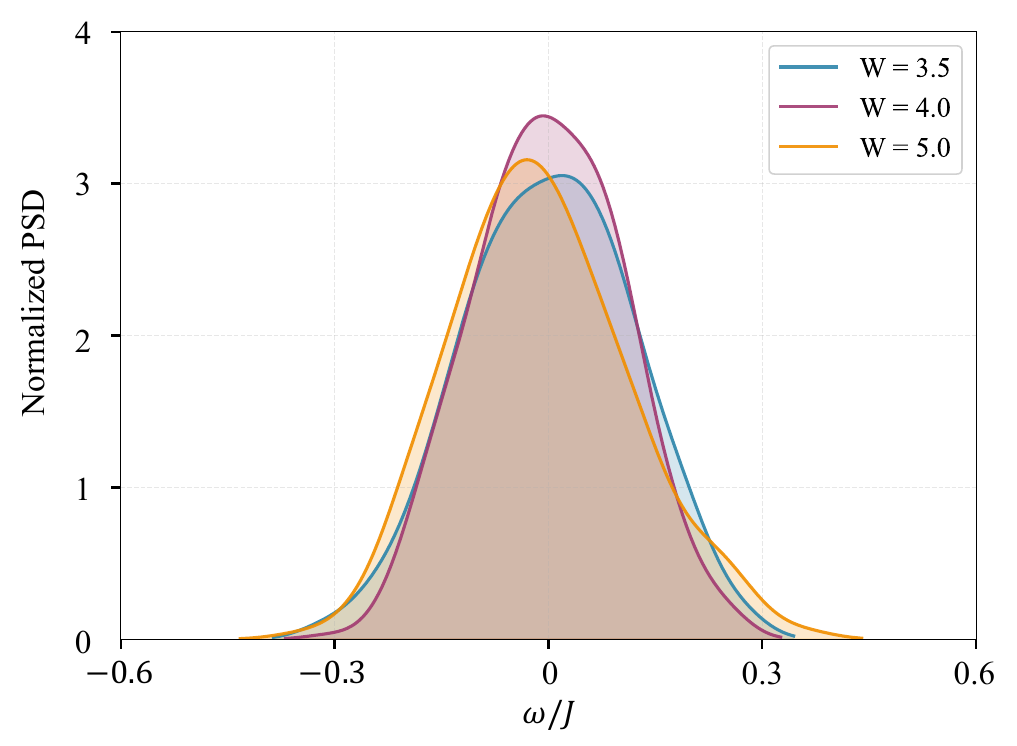}
\caption{Average PSD for disorder strengths $W=3.5$, $4.0$, and $5.0$. For each value of $W$, the PSD is averaged over 100 disorder realizations. The resulting distribution is then smoothed using Gaussian kernel density estimation (KDE) and normalized. All calculations are performed with $u=-2.1$.}
\label{figap:6}
\end{figure}

\subsection{Local perturbation and global disorder reconfiguration}
We examine the system’s robustness against these perturbations—including defects and locally reconstructed disorder—by evaluating both the single-shot emission and the changes in the realization-averaged PSD. The single mode emission is guaranteed by topological protection, while the latter corresponds to an eigenvalue sensitivity problem. For small perturbations of the array, the response can be analyzed using first-order perturbation theory~\cite{Sakurai:2011zzS}. That is, under a perturbation $H \to H + H_{\text{perturbation}}$, the change of the eigenstates is given by:
\begin{equation}
    \label{eq:5}
    \delta \omega_j = \bra {v^{(j)}}H_{\text{perturbation}}\ket {v^{(j)}}.
\end{equation}

Since the perturbation Hamiltonian is sparse—having nonzero matrix elements only on the defect site and its neighboring sites for defects, or only on the reconstructed sites for disorder reconfiguration—the magnitude of the eigenstate variation is essentially determined by the overlap between the perturbation and the unperturbed eigenstate $\delta \omega_j \sim \sum_{m,n\text{ in changed sites}} |(v^{(j)})_{m,n}|^2J$. Considering our $20 \times 20$ system, and assuming that the mode is approximately uniformly distributed over the outer two boundary layers comprising about $140$ sites, while fewer than ten boundary sites are modified, the resulting perturbation to the lasing frequency is of order $(10^{-2})J$. In practice, however, the mode profile in a TARL is highly nonuniform, and the actual frequency shift depends on the specific locations of the modified sites. Many sites, including both bulk sites and portions of the boundary, carry negligible weight of the edge-state wave function and are therefore insensitive to local perturbations. Consequently, TARLs exhibit strong robustness against localized disturbances.

On the other hand, if the global disorder configuration is reconstructed, as shown in Fig.~\ref{figap:6}, the averaged PSD spreads across the entire mobility gap. Since the mobility gap is relatively narrow, the resulting PSD remains narrower than the bandwidth of the gain profile. It is worth noting that this distribution does not correspond to the actual emission spectrum observed in experiments. Instead, it reflects the robustness of the system against strong perturbations.

\subsection{Temporal coherence}
%This behavior is more similar to the coherence properties of a single-mode laser rather than the KPZ-type behavior observed in topological lasers~\cite{fontaineKardar2022, amelioTheoryCoherenceTopological2020}.
Previous study~\cite{amelioTheoryCoherenceTopological2020S} has shown that the edge states of topological lasers exhibit a linear dispersion relation, which leads to pronounced oscillations in their coherence. As a result, it becomes difficult to directly extract the intrinsic coherence decay behavior. To address this issue, a transformation to the comoving frame was introduced. In contrast, the edge states of TARLs do not possess a well-defined dispersion relation  and can rapidly evolve into a single-mode-dominated steady state. However, repeated realizations may reach steady states with slightly different frequencies, which still induce oscillations in the coherence. Therefore, in a similar manner, we eliminate the evolution associated with the eigenfrequency from the mode coefficients:
\begin{equation}
\label{eq:9}
{c}(t) = A \exp[-i\phi(t)]
\end{equation}
This allows us to express the coherence function as:
\begin{equation}\label{eq:10}
g^{(1)}(t) = \frac{|\langle c^*(0)c(t)\rangle_\text{ens}|}{\sqrt{\langle c^*(0)c(0)\rangle_\text{ens}\langle c^*(t)c(t)\rangle_\text{ens}}}.
\end{equation}
This formulation provides a framework for analyzing the coherence properties of TARLs.

\subsection{Derivation of the phase equation}

Our derivation starts from the dynamical equation
\begin{equation}\label{eq11}
    i\partial_t \Psi_{m,n} = (H\Psi)_{m,n} 
+ i\left( \frac{P_{m,n}(\omega)}{1 + \beta \langle\Psi|\Psi\rangle} - \gamma \right)\Psi_{m,n} 
+ \sqrt{2D_{m,n}}\,\xi_{m,n},
\end{equation}
together with the single-mode ansatz, which leads to
\begin{equation}\label{eq12}
\Psi_{m,n}(t) =
    A(t) v^{(j)}_{m,n} e^{-i\omega_jt-i\phi(t)} + \sum_{j'\neq j} \delta c_{j'}(t)e^{-i\omega_{j'}t}v^{(j')}_{m,n}.
\end{equation}
Substituting Eq.~\ref{eq12} into Eq.~\ref{eq11} yields
\begin{equation}\label{eq13}
\begin{aligned}
&\dot A v^{(j)}_{m,n} e^{-i\omega_j t-i\phi}
- i\dot\phi A v^{(j)}_{m,n} e^{-i\omega_j t-i\phi}
+ \sum_{j'\ne j} \delta \dot c_{j'} v^{(j')}_{m,n}e^{-i\omega_{j'}t}\\
%= -i\sum_{j'\ne j} \omega_{j'} \delta c_{j'} v^{(j')}_{m,n} \\
&= \left( \frac{\mathcal{P}(\omega)G_{m,n}}{1 + \beta \langle\Psi|\Psi\rangle} - \gamma \right)
\left(A(t) v^{(j)}_{m,n} e^{-i\omega_j t-i\phi(t)}
+ \sum_{j'\neq j} \delta c_{j'}(t)v^{(j')}_{m,n}\right)+\sqrt{2D_{m,n}}\,\xi_{m,n},
\end{aligned}
\end{equation}
where the gain is decomposed into a frequency-dependent amplitude $\mathcal P(\omega)$ and a spatial profile $G_{m,n}$. We focus on the steady-state evolution. Therefore, for the states within the gain window, we approximate $\frac{\mathcal P(\omega)}{1+\beta \langle\Psi|\Psi\rangle}=\gamma$. 

Thus, based on the orthogonality, we obtain
\begin{equation}\label{eq15}
\begin{aligned}
\dot A e^{-i\omega_j t-i\phi}-i\dot \phi A e^{-i\omega_j t-i\phi}=& \gamma (\bra{v^{(j)}}G\ket{v^{(j)}}-1)Ae^{-i\omega_j t-i\phi} +\gamma \sum_{j'\ne j} \bra{v^{(j)}}G\ket{v^{(j')}}\delta c_{j'} \\&+\sqrt{2\gamma}\xi_j,
\end{aligned}
\end{equation}
where $\xi_j =\sum_{m,n} \bra{v^{(j)}_{m,n}}\sqrt{2D_{m,n}}|\xi_{m,n}\rangle/\sqrt{2\gamma}$. It satisfies $\langle\xi_j^*(t') \xi_j(t) \rangle \simeq \delta(t-t')$ indicating that it is essentially white noise. Meanwhile, $\bra{v^{(j)}}G\ket{v^{(j)}}-1$ is typically small.

Separating the real and imaginary parts of Eq.~\ref{eq15} we obtain
\begin{equation}\label{eq17}
    \dot A = \gamma (\bra{v^{(j)}}G\ket{v^{(j)}}-1)A +\gamma \sum_{j'\ne j} \text{Re}\bra{v^{(j)}}G\ket{v^{(j')}}\frac{\delta c_{j'}e^{-i\omega_{j'}t }}{ e^{-i\omega_j t-i\phi}} +\sqrt{2\gamma}\xi_A,
\end{equation}
\begin{equation}\label{eq18}
    \dot \phi = \frac{\sqrt {2\gamma}}{A} \xi_\phi -\gamma \sum_{j'\ne j} \text{Im} \left[\langle v^{(j)}|G|v^{(j')}\rangle \frac{\delta c_{j'}e^{-i\omega_{j'}t }}{Ae^{-i\omega_j t-i\phi}}\right],
\end{equation}
where $\xi_A$ and $\xi_\phi$ denote the real and imaginary parts of the noise $\xi_j$, respectively.

Eq.~\ref{eq18} is the phase evolution equation presented in the main text.

\subsection{Haldane-like model}

\begin{figure}
\centering
\includegraphics[width=\linewidth]{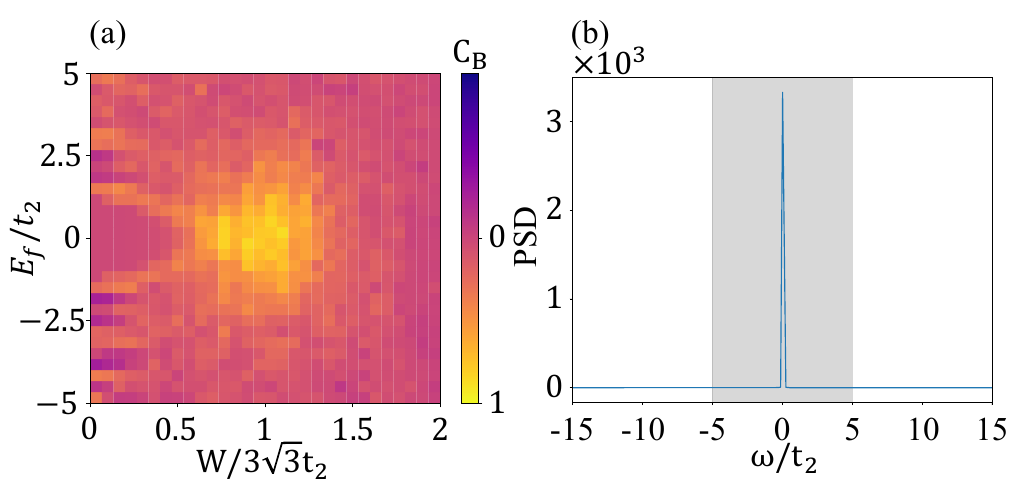}
\caption{Results for the Haldane model (a) Disorder-induced topological phase transition in the Haldane-like model. Evolution of the energy gap as a function of disorder strength $W$. (b) PSD of a single-shot single-mode lasing event, where the gray region denotes the gain-frequency window. Model parameters: $t_{1}=1$, $t_{2}=0.2$, $\theta=\pi/2$, $M = 1.05\times 3\sqrt{3}\,t_{2}$, $P=0.5$, $\gamma = 0.01$ and $\beta = 1$. The system size is $20\times20$ with open boundaries in the zigzag geometry.}
\label{figap:1} 
\end{figure}

\begin{figure*}
\centering
\includegraphics[width=\linewidth]{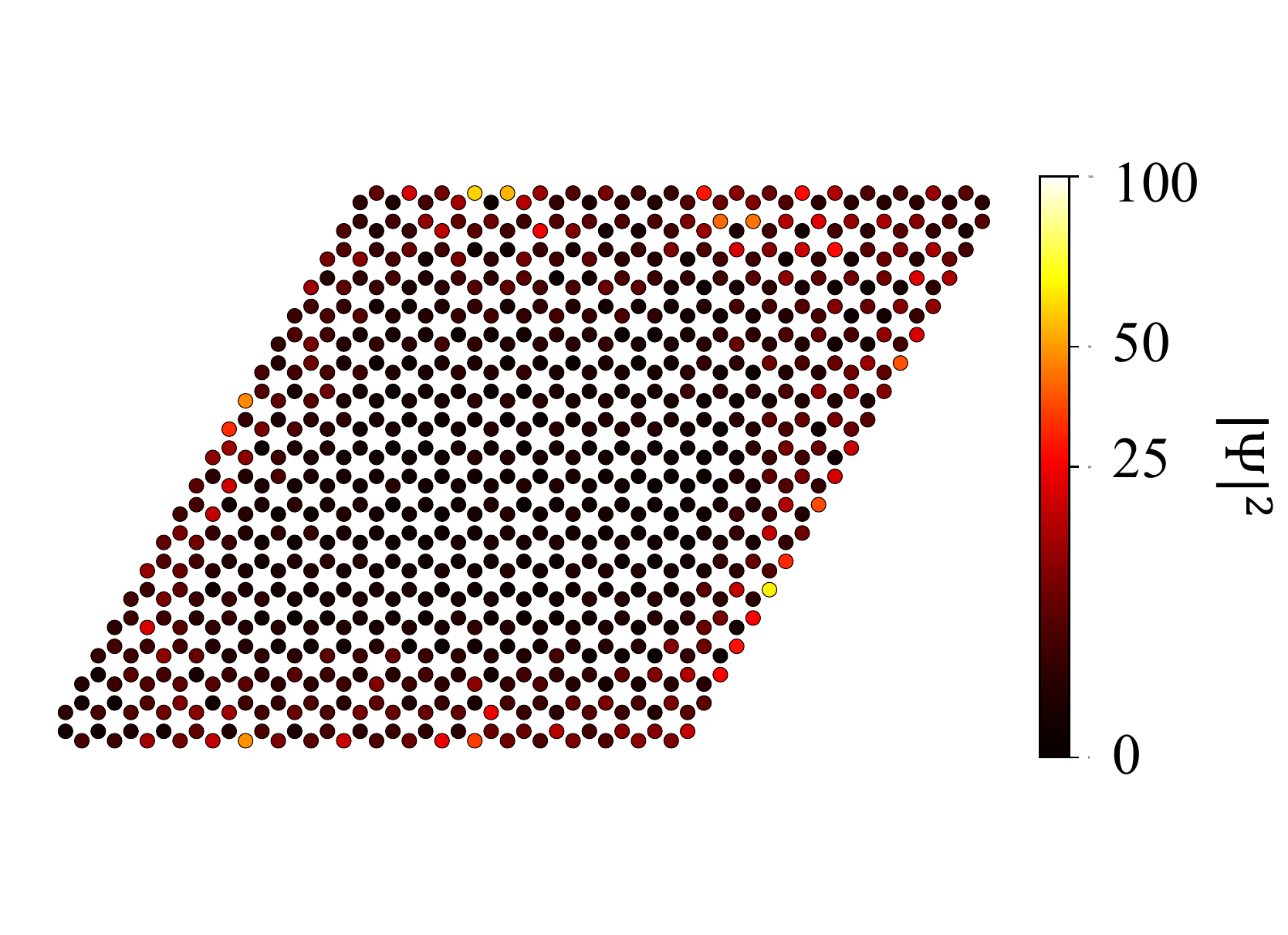}
\caption{Single-mode lasing pattern in the Haldane-based TARL of Fig.~\ref{figap:1} (b). The parameters are the same as that in Fig.~\ref{figap:1}.}
\label{figap:2} 
\end{figure*}

In the preceding sections, we analyzed a representative minimal TARL model and examined its behavior with and without noise. However, experimental realizations of photonic topological Anderson insulator are typically implemented in systems whose topological phase diagram resembles that of the Haldane model~\cite{stutzerPhotonicTopologicalAnderson2018S, liuTopologicalAndersonInsulator2020S, chenRealizationTimeReversalInvariant2024S, cuiPhotonic2Topological2022S}. For this reason, we further construct a TARL based on the Haldane model. The Hamiltonian of the Haldane model is given by:

\begin{equation}
\label{eq:14}
\begin{split}
H ={}& -t_1 \sum_{\langle m,n\rangle} \psi_m^\dagger \psi_n
      - t_{2} \sum_{\langle\!\langle m,n\rangle\!\rangle}
      e^{i \mu_{mn}\theta}\, \psi_m^\dagger \psi_n \\
    & + M \sum_{m} \kappa_m\, \psi_m^\dagger \psi_m.
\end{split}
\end{equation}

Here, $\langle m,n\rangle (\langle\!\langle m,n\rangle\!\rangle)$ denotes a sum over nearest-neighbor (next-nearest-neighbor) site pairs. In this case,  $\psi$ no longer takes the form of a spinor. $\psi_m^\dagger \psi_n$ is the hopping operator from site $n$ to site $m$, and $t_1 (t_{2})$ is the nearest- (next-nearest-) neighbor hopping amplitude. The sign $\mu_{mn}=\pm1$ assigns the orientation of the next-nearest-neighbor hopping phase: $\mu_{mn}=+1 (-1)$ when the hopping from $n$ to $m$ corresponds to positive (negative) circulation around the hexagonal plaquette (equivalently, $\mu_{mn}=\operatorname{sgn}[(\mathbf{d}_{mk}\times\mathbf{d}_{kn})_{z}]$ with $k$ the intermediate nearest-neighbor site). $M$ denotes the sublattice mass, and $\kappa_m=\pm1$ labels sites on the two sublattices.

The clean system’s topology is controlled by the ratio $M/t_{2}$ and the phase $\theta$: the bulk gap closes at $|M/t_2|=\pm 3\sqrt{3}\sin\theta$, and the system is in a topological phase (Chern number $|C|=1$) when
$|M/t_2| < 3\sqrt{3}|\sin\theta|$.
If the clean system is tuned to the trivial regime, $|M/t_2| > 3\sqrt{3}|\sin\theta|$, and we add an on-site potential drawn independently from a uniform distribution $[ -W/2, W/2]$ (where $W$ denotes the disorder strength), the system can be driven into a TAI phase for an appropriate range of $W$. 

We consider a $20 \times 20$ system with open boundaries in the zigzag geometry. 
The model parameters are chosen as 
$t_{1}=1$, $t_{2}=0.2$, $\theta=\pi/2$, and 
$M = 1.05 \times 3\sqrt{3}\, t_{2}$, 
which place the corresponding clean system in the trivial regime. Upon introducing on-site disorder, we monitor the evolution of the band gap as the disorder strength W increases. As shown in Fig. \ref{figap:1}(a), the trivial bulk gap gradually closes with increasing W; at larger disorder strengths, a new gap reopens, indicating that the system has entered the TAI phase. 

To examine whether single-mode TARL emission can be realized in this model, 
We introduce onsite disorder of strength 
$W = 1.4 \times 3\sqrt{3}\, t_{2}$ 
and apply gain along the boundary over a frequency window 
$[-5t_{2},\, 5t_{2}]$, following the same procedure used in the model discussed earlier. The gain parameters are chosen as $\gamma = 0.01$, $P = 0.5$, and $\beta = 1$. Under these conditions, we observe single-mode edge-state lasing, indicating that TARL operation is achievable not only in minimal theoretical models but also in more experimentally relevant Haldane-type systems. [Fig.~\ref{figap:1}(b), Fig.~\ref{figap:2}]

% Bibliography
%\clearpage
%\bibliography{citation}

\begin{thebibliography}{50}%
\makeatletter
\providecommand \@ifxundefined [1]{%
 \@ifx{#1\undefined}
}%
\providecommand \@ifnum [1]{%
 \ifnum #1\expandafter \@firstoftwo
 \else \expandafter \@secondoftwo
 \fi
}%
\providecommand \@ifx [1]{%
 \ifx #1\expandafter \@firstoftwo
 \else \expandafter \@secondoftwo
 \fi
}%
\providecommand \natexlab [1]{#1}%
\providecommand \enquote  [1]{``#1''}%
\providecommand \bibnamefont  [1]{#1}%
\providecommand \bibfnamefont [1]{#1}%
\providecommand \citenamefont [1]{#1}%
\providecommand \href@noop [0]{\@secondoftwo}%
\providecommand \href [0]{\begingroup \@sanitize@url \@href}%
\providecommand \@href[1]{\@@startlink{#1}\@@href}%
\providecommand \@@href[1]{\endgroup#1\@@endlink}%
\providecommand \@sanitize@url [0]{\catcode `\\12\catcode `\$12\catcode
  `\&12\catcode `\#12\catcode `\^12\catcode `\_12\catcode `\%12\relax}%
\providecommand \@@startlink[1]{}%
\providecommand \@@endlink[0]{}%
\providecommand \url  [0]{\begingroup\@sanitize@url \@url }%
\providecommand \@url [1]{\endgroup\@href {#1}{\urlprefix }}%
\providecommand \urlprefix  [0]{URL }%
\providecommand \Eprint [0]{\href }%
\providecommand \doibase [0]{https://doi.org/}%
\providecommand \selectlanguage [0]{\@gobble}%
\providecommand \bibinfo  [0]{\@secondoftwo}%
\providecommand \bibfield  [0]{\@secondoftwo}%
\providecommand \translation [1]{[#1]}%
\providecommand \BibitemOpen [0]{}%
\providecommand \bibitemStop [0]{}%
\providecommand \bibitemNoStop [0]{.\EOS\space}%
\providecommand \EOS [0]{\spacefactor3000\relax}%
\providecommand \BibitemShut  [1]{\csname bibitem#1\endcsname}%
\let\auto@bib@innerbib\@empty
%</preamble>
\bibitem [{\citenamefont {Abbott}\ \emph {et~al.}(2016)\citenamefont {Abbott},
  \citenamefont {Abbott}, \citenamefont {Abbott}, \citenamefont {Abernathy},
  \citenamefont {Acernese}, \citenamefont {Ackley}, \citenamefont {Adams},
  \citenamefont {Adams}, \citenamefont {Addesso}, \citenamefont {Adhikari}
  \emph {et~al.}}]{abbott2016observation}%
  \BibitemOpen
  \bibfield  {author} {\bibinfo {author} {\bibfnamefont {B.~P.}\ \bibnamefont
  {Abbott}}, \bibinfo {author} {\bibfnamefont {R.}~\bibnamefont {Abbott}},
  \bibinfo {author} {\bibfnamefont {T.}~\bibnamefont {Abbott}}, \bibinfo
  {author} {\bibfnamefont {M.}~\bibnamefont {Abernathy}}, \bibinfo {author}
  {\bibfnamefont {F.}~\bibnamefont {Acernese}}, \bibinfo {author}
  {\bibfnamefont {K.}~\bibnamefont {Ackley}}, \bibinfo {author} {\bibfnamefont
  {C.}~\bibnamefont {Adams}}, \bibinfo {author} {\bibfnamefont
  {T.}~\bibnamefont {Adams}}, \bibinfo {author} {\bibfnamefont
  {P.}~\bibnamefont {Addesso}}, \bibinfo {author} {\bibfnamefont
  {R.}~\bibnamefont {Adhikari}}, \emph {et~al.},\ }\href
  {https://doi.org/10.1103/PhysRevLett.116.061102} {\bibfield  {journal}
  {\bibinfo  {journal} {Physical Review Letters}\ }\textbf {\bibinfo {volume}
  {116}},\ \bibinfo {pages} {061102} (\bibinfo {year} {2016})}\BibitemShut
  {NoStop}%
\bibitem [{\citenamefont {Hell}(2007)}]{hell2007far}%
  \BibitemOpen
  \bibfield  {author} {\bibinfo {author} {\bibfnamefont {S.~W.}\ \bibnamefont
  {Hell}},\ }\href {https://doi.org/10.1126/science.1137395} {\bibfield
  {journal} {\bibinfo  {journal} {Science}\ }\textbf {\bibinfo {volume}
  {316}},\ \bibinfo {pages} {1153} (\bibinfo {year} {2007})}\BibitemShut
  {NoStop}%
\bibitem [{\citenamefont {Xu}\ \emph {et~al.}(2021)\citenamefont {Xu},
  \citenamefont {Liu}, \citenamefont {Wan}, \citenamefont {Li}, \citenamefont
  {Huang}, \citenamefont {Guo}, \citenamefont {Ma},\ and\ \citenamefont
  {Pan}}]{xu2021non}%
  \BibitemOpen
  \bibfield  {author} {\bibinfo {author} {\bibfnamefont {F.}~\bibnamefont
  {Xu}}, \bibinfo {author} {\bibfnamefont {B.}~\bibnamefont {Liu}}, \bibinfo
  {author} {\bibfnamefont {Z.}~\bibnamefont {Wan}}, \bibinfo {author}
  {\bibfnamefont {X.}~\bibnamefont {Li}}, \bibinfo {author} {\bibfnamefont
  {G.}~\bibnamefont {Huang}}, \bibinfo {author} {\bibfnamefont
  {H.}~\bibnamefont {Guo}}, \bibinfo {author} {\bibfnamefont {X.}~\bibnamefont
  {Ma}},\ and\ \bibinfo {author} {\bibfnamefont {J.-W.}\ \bibnamefont {Pan}},\
  }\href {https://doi.org/10.1103/PhysRevLett.127.053602} {\bibfield  {journal}
  {\bibinfo  {journal} {Physical Review Letters}\ }\textbf {\bibinfo {volume}
  {127}},\ \bibinfo {pages} {053602} (\bibinfo {year} {2021})}\BibitemShut
  {NoStop}%
\bibitem [{\citenamefont {Marin-Palomo}\ \emph {et~al.}(2017)\citenamefont
  {Marin-Palomo}, \citenamefont {Kemal}, \citenamefont {Karpov}, \citenamefont
  {Kordts}, \citenamefont {Pfeifle}, \citenamefont {Pfeiffer}, \citenamefont
  {Trocha}, \citenamefont {Wolf}, \citenamefont {Brasch}, \citenamefont
  {Anderson} \emph {et~al.}}]{marin2017microresonator}%
  \BibitemOpen
  \bibfield  {author} {\bibinfo {author} {\bibfnamefont {P.}~\bibnamefont
  {Marin-Palomo}}, \bibinfo {author} {\bibfnamefont {J.~N.}\ \bibnamefont
  {Kemal}}, \bibinfo {author} {\bibfnamefont {M.}~\bibnamefont {Karpov}},
  \bibinfo {author} {\bibfnamefont {A.}~\bibnamefont {Kordts}}, \bibinfo
  {author} {\bibfnamefont {J.}~\bibnamefont {Pfeifle}}, \bibinfo {author}
  {\bibfnamefont {M.~H.}\ \bibnamefont {Pfeiffer}}, \bibinfo {author}
  {\bibfnamefont {P.}~\bibnamefont {Trocha}}, \bibinfo {author} {\bibfnamefont
  {S.}~\bibnamefont {Wolf}}, \bibinfo {author} {\bibfnamefont {V.}~\bibnamefont
  {Brasch}}, \bibinfo {author} {\bibfnamefont {M.~H.}\ \bibnamefont
  {Anderson}}, \emph {et~al.},\ }\href {https://doi.org/10.1038/nature22387}
  {\bibfield  {journal} {\bibinfo  {journal} {Nature}\ }\textbf {\bibinfo
  {volume} {546}},\ \bibinfo {pages} {274} (\bibinfo {year}
  {2017})}\BibitemShut {NoStop}%
\bibitem [{\citenamefont {Zhong}\ \emph {et~al.}(2020)\citenamefont {Zhong},
  \citenamefont {Wang}, \citenamefont {Deng}, \citenamefont {Chen},
  \citenamefont {Peng}, \citenamefont {Luo}, \citenamefont {Qin}, \citenamefont
  {Wu}, \citenamefont {Ding}, \citenamefont {Hu} \emph
  {et~al.}}]{zhong2020quantum}%
  \BibitemOpen
  \bibfield  {author} {\bibinfo {author} {\bibfnamefont {H.-S.}\ \bibnamefont
  {Zhong}}, \bibinfo {author} {\bibfnamefont {H.}~\bibnamefont {Wang}},
  \bibinfo {author} {\bibfnamefont {Y.-H.}\ \bibnamefont {Deng}}, \bibinfo
  {author} {\bibfnamefont {M.-C.}\ \bibnamefont {Chen}}, \bibinfo {author}
  {\bibfnamefont {L.-C.}\ \bibnamefont {Peng}}, \bibinfo {author}
  {\bibfnamefont {Y.-H.}\ \bibnamefont {Luo}}, \bibinfo {author} {\bibfnamefont
  {J.}~\bibnamefont {Qin}}, \bibinfo {author} {\bibfnamefont {D.}~\bibnamefont
  {Wu}}, \bibinfo {author} {\bibfnamefont {X.}~\bibnamefont {Ding}}, \bibinfo
  {author} {\bibfnamefont {Y.}~\bibnamefont {Hu}}, \emph {et~al.},\ }\href
  {https://doi.org/10.1126/science.abe8770} {\bibfield  {journal} {\bibinfo
  {journal} {Science}\ }\textbf {\bibinfo {volume} {370}},\ \bibinfo {pages}
  {1460} (\bibinfo {year} {2020})}\BibitemShut {NoStop}%
\bibitem [{\citenamefont {Carolan}\ \emph {et~al.}(2015)\citenamefont
  {Carolan}, \citenamefont {Harrold}, \citenamefont {Sparrow}, \citenamefont
  {Mart{\'\i}n-L{\'o}pez}, \citenamefont {Russell}, \citenamefont
  {Silverstone}, \citenamefont {Shadbolt}, \citenamefont {Matsuda},
  \citenamefont {Oguma}, \citenamefont {Itoh} \emph
  {et~al.}}]{carolan2015universal}%
  \BibitemOpen
  \bibfield  {author} {\bibinfo {author} {\bibfnamefont {J.}~\bibnamefont
  {Carolan}}, \bibinfo {author} {\bibfnamefont {C.}~\bibnamefont {Harrold}},
  \bibinfo {author} {\bibfnamefont {C.}~\bibnamefont {Sparrow}}, \bibinfo
  {author} {\bibfnamefont {E.}~\bibnamefont {Mart{\'\i}n-L{\'o}pez}}, \bibinfo
  {author} {\bibfnamefont {N.~J.}\ \bibnamefont {Russell}}, \bibinfo {author}
  {\bibfnamefont {J.~W.}\ \bibnamefont {Silverstone}}, \bibinfo {author}
  {\bibfnamefont {P.~J.}\ \bibnamefont {Shadbolt}}, \bibinfo {author}
  {\bibfnamefont {N.}~\bibnamefont {Matsuda}}, \bibinfo {author} {\bibfnamefont
  {M.}~\bibnamefont {Oguma}}, \bibinfo {author} {\bibfnamefont
  {M.}~\bibnamefont {Itoh}}, \emph {et~al.},\ }\href
  {https://doi.org/10.1126/science.aab3642} {\bibfield  {journal} {\bibinfo
  {journal} {Science}\ }\textbf {\bibinfo {volume} {349}},\ \bibinfo {pages}
  {711} (\bibinfo {year} {2015})}\BibitemShut {NoStop}%
\bibitem [{\citenamefont {Thyagarajan}\ and\ \citenamefont
  {Ghatak}(2011)}]{Thyagarajan2011}%
  \BibitemOpen
  \bibfield  {author} {\bibinfo {author} {\bibfnamefont {K.}~\bibnamefont
  {Thyagarajan}}\ and\ \bibinfo {author} {\bibfnamefont {A.}~\bibnamefont
  {Ghatak}},\ }\bibinfo {title} {Semiclassical theory of the laser},\ in\ \href
  {https://doi.org/10.1007/978-1-4419-6442-7_6} {\emph {\bibinfo {booktitle}
  {Lasers: Fundamentals and Applications}}}\ (\bibinfo  {publisher} {Springer
  US},\ \bibinfo {address} {Boston, MA},\ \bibinfo {year} {2011})\ pp.\
  \bibinfo {pages} {121--141}\BibitemShut {NoStop}%
\bibitem [{\citenamefont {Hentschel}\ \emph {et~al.}(2001)\citenamefont
  {Hentschel}, \citenamefont {Kienberger}, \citenamefont {Spielmann},
  \citenamefont {Reider}, \citenamefont {Milosevic}, \citenamefont {Brabec},
  \citenamefont {Corkum}, \citenamefont {Heinzmann}, \citenamefont {Drescher},\
  and\ \citenamefont {Krausz}}]{hentschel2001attosecond}%
  \BibitemOpen
  \bibfield  {author} {\bibinfo {author} {\bibfnamefont {M.}~\bibnamefont
  {Hentschel}}, \bibinfo {author} {\bibfnamefont {R.}~\bibnamefont
  {Kienberger}}, \bibinfo {author} {\bibfnamefont {C.}~\bibnamefont
  {Spielmann}}, \bibinfo {author} {\bibfnamefont {G.~A.}\ \bibnamefont
  {Reider}}, \bibinfo {author} {\bibfnamefont {N.}~\bibnamefont {Milosevic}},
  \bibinfo {author} {\bibfnamefont {T.}~\bibnamefont {Brabec}}, \bibinfo
  {author} {\bibfnamefont {P.}~\bibnamefont {Corkum}}, \bibinfo {author}
  {\bibfnamefont {U.}~\bibnamefont {Heinzmann}}, \bibinfo {author}
  {\bibfnamefont {M.}~\bibnamefont {Drescher}},\ and\ \bibinfo {author}
  {\bibfnamefont {F.}~\bibnamefont {Krausz}},\ }\href
  {https://doi.org/10.1038/35107000} {\bibfield  {journal} {\bibinfo  {journal}
  {Nature}\ }\textbf {\bibinfo {volume} {414}},\ \bibinfo {pages} {509}
  (\bibinfo {year} {2001})}\BibitemShut {NoStop}%
\bibitem [{\citenamefont {Diddams}\ \emph {et~al.}(2000)\citenamefont
  {Diddams}, \citenamefont {Jones}, \citenamefont {Ye}, \citenamefont
  {Cundiff}, \citenamefont {Hall}, \citenamefont {Ranka}, \citenamefont
  {Windeler}, \citenamefont {Holzwarth}, \citenamefont {Udem},\ and\
  \citenamefont {H{\"a}nsch}}]{diddams2000direct}%
  \BibitemOpen
  \bibfield  {author} {\bibinfo {author} {\bibfnamefont {S.~A.}\ \bibnamefont
  {Diddams}}, \bibinfo {author} {\bibfnamefont {D.~J.}\ \bibnamefont {Jones}},
  \bibinfo {author} {\bibfnamefont {J.}~\bibnamefont {Ye}}, \bibinfo {author}
  {\bibfnamefont {S.~T.}\ \bibnamefont {Cundiff}}, \bibinfo {author}
  {\bibfnamefont {J.~L.}\ \bibnamefont {Hall}}, \bibinfo {author}
  {\bibfnamefont {J.~K.}\ \bibnamefont {Ranka}}, \bibinfo {author}
  {\bibfnamefont {R.~S.}\ \bibnamefont {Windeler}}, \bibinfo {author}
  {\bibfnamefont {R.}~\bibnamefont {Holzwarth}}, \bibinfo {author}
  {\bibfnamefont {T.}~\bibnamefont {Udem}},\ and\ \bibinfo {author}
  {\bibfnamefont {T.~W.}\ \bibnamefont {H{\"a}nsch}},\ }\href
  {https://doi.org/10.1103/PhysRevLett.84.5102} {\bibfield  {journal} {\bibinfo
   {journal} {Physical Review Letters}\ }\textbf {\bibinfo {volume} {84}},\
  \bibinfo {pages} {5102} (\bibinfo {year} {2000})}\BibitemShut {NoStop}%
\bibitem [{\citenamefont {St-Jean}\ \emph {et~al.}(2017)\citenamefont
  {St-Jean}, \citenamefont {Goblot}, \citenamefont {Galopin}, \citenamefont
  {Lema{\^\i}tre}, \citenamefont {Ozawa}, \citenamefont {Le~Gratiet},
  \citenamefont {Sagnes}, \citenamefont {Bloch},\ and\ \citenamefont
  {Amo}}]{st2017lasing}%
  \BibitemOpen
  \bibfield  {author} {\bibinfo {author} {\bibfnamefont {P.}~\bibnamefont
  {St-Jean}}, \bibinfo {author} {\bibfnamefont {V.}~\bibnamefont {Goblot}},
  \bibinfo {author} {\bibfnamefont {E.}~\bibnamefont {Galopin}}, \bibinfo
  {author} {\bibfnamefont {A.}~\bibnamefont {Lema{\^\i}tre}}, \bibinfo {author}
  {\bibfnamefont {T.}~\bibnamefont {Ozawa}}, \bibinfo {author} {\bibfnamefont
  {L.}~\bibnamefont {Le~Gratiet}}, \bibinfo {author} {\bibfnamefont
  {I.}~\bibnamefont {Sagnes}}, \bibinfo {author} {\bibfnamefont
  {J.}~\bibnamefont {Bloch}},\ and\ \bibinfo {author} {\bibfnamefont
  {A.}~\bibnamefont {Amo}},\ }\href {https://doi.org/10.1038/s41566-017-0006-2}
  {\bibfield  {journal} {\bibinfo  {journal} {Nature Photonics}\ }\textbf
  {\bibinfo {volume} {11}},\ \bibinfo {pages} {651} (\bibinfo {year}
  {2017})}\BibitemShut {NoStop}%
\bibitem [{\citenamefont {Bahari}\ \emph {et~al.}(2017)\citenamefont {Bahari},
  \citenamefont {Ndao}, \citenamefont {Vallini}, \citenamefont {Amili},
  \citenamefont {Fainman},\ and\ \citenamefont
  {Kanté}}]{bahari2017nonreciprocal}%
  \BibitemOpen
  \bibfield  {author} {\bibinfo {author} {\bibfnamefont {B.}~\bibnamefont
  {Bahari}}, \bibinfo {author} {\bibfnamefont {A.}~\bibnamefont {Ndao}},
  \bibinfo {author} {\bibfnamefont {F.}~\bibnamefont {Vallini}}, \bibinfo
  {author} {\bibfnamefont {A.~E.}\ \bibnamefont {Amili}}, \bibinfo {author}
  {\bibfnamefont {Y.}~\bibnamefont {Fainman}},\ and\ \bibinfo {author}
  {\bibfnamefont {B.}~\bibnamefont {Kanté}},\ }\href
  {https://doi.org/10.1126/science.aao4551} {\bibfield  {journal} {\bibinfo
  {journal} {Science}\ }\textbf {\bibinfo {volume} {358}},\ \bibinfo {pages}
  {636} (\bibinfo {year} {2017})}\BibitemShut {NoStop}%
\bibitem [{\citenamefont {Harari}\ \emph {et~al.}(2018)\citenamefont {Harari},
  \citenamefont {Bandres}, \citenamefont {Lumer}, \citenamefont {Rechtsman},
  \citenamefont {Chong}, \citenamefont {Khajavikhan}, \citenamefont
  {Christodoulides},\ and\ \citenamefont {Segev}}]{harari2018topological}%
  \BibitemOpen
  \bibfield  {author} {\bibinfo {author} {\bibfnamefont {G.}~\bibnamefont
  {Harari}}, \bibinfo {author} {\bibfnamefont {M.~A.}\ \bibnamefont {Bandres}},
  \bibinfo {author} {\bibfnamefont {Y.}~\bibnamefont {Lumer}}, \bibinfo
  {author} {\bibfnamefont {M.~C.}\ \bibnamefont {Rechtsman}}, \bibinfo {author}
  {\bibfnamefont {Y.~D.}\ \bibnamefont {Chong}}, \bibinfo {author}
  {\bibfnamefont {M.}~\bibnamefont {Khajavikhan}}, \bibinfo {author}
  {\bibfnamefont {D.~N.}\ \bibnamefont {Christodoulides}},\ and\ \bibinfo
  {author} {\bibfnamefont {M.}~\bibnamefont {Segev}},\ }\href
  {https://doi.org/10.1126/science.aar4003} {\bibfield  {journal} {\bibinfo
  {journal} {Science}\ }\textbf {\bibinfo {volume} {359}},\ \bibinfo {pages}
  {eaar4003} (\bibinfo {year} {2018})}\BibitemShut {NoStop}%
\bibitem [{\citenamefont {Bandres}\ \emph {et~al.}(2018)\citenamefont
  {Bandres}, \citenamefont {Wittek}, \citenamefont {Harari}, \citenamefont
  {Parto}, \citenamefont {Ren}, \citenamefont {Segev}, \citenamefont
  {Christodoulides},\ and\ \citenamefont
  {Khajavikhan}}]{bandres2018topological}%
  \BibitemOpen
  \bibfield  {author} {\bibinfo {author} {\bibfnamefont {M.~A.}\ \bibnamefont
  {Bandres}}, \bibinfo {author} {\bibfnamefont {S.}~\bibnamefont {Wittek}},
  \bibinfo {author} {\bibfnamefont {G.}~\bibnamefont {Harari}}, \bibinfo
  {author} {\bibfnamefont {M.}~\bibnamefont {Parto}}, \bibinfo {author}
  {\bibfnamefont {J.}~\bibnamefont {Ren}}, \bibinfo {author} {\bibfnamefont
  {M.}~\bibnamefont {Segev}}, \bibinfo {author} {\bibfnamefont {D.~N.}\
  \bibnamefont {Christodoulides}},\ and\ \bibinfo {author} {\bibfnamefont
  {M.}~\bibnamefont {Khajavikhan}},\ }\href
  {https://doi.org/10.1126/science.aar4005} {\bibfield  {journal} {\bibinfo
  {journal} {Science}\ }\textbf {\bibinfo {volume} {359}},\ \bibinfo {pages}
  {eaar4005} (\bibinfo {year} {2018})}\BibitemShut {NoStop}%
\bibitem [{\citenamefont {Ozawa}\ \emph {et~al.}(2019)\citenamefont {Ozawa},
  \citenamefont {Price}, \citenamefont {Amo}, \citenamefont {Goldman},
  \citenamefont {Hafezi}, \citenamefont {Lu}, \citenamefont {Rechtsman},
  \citenamefont {Schuster}, \citenamefont {Simon}, \citenamefont {Zilberberg}
  \emph {et~al.}}]{ozawa2019topological}%
  \BibitemOpen
  \bibfield  {author} {\bibinfo {author} {\bibfnamefont {T.}~\bibnamefont
  {Ozawa}}, \bibinfo {author} {\bibfnamefont {H.~M.}\ \bibnamefont {Price}},
  \bibinfo {author} {\bibfnamefont {A.}~\bibnamefont {Amo}}, \bibinfo {author}
  {\bibfnamefont {N.}~\bibnamefont {Goldman}}, \bibinfo {author} {\bibfnamefont
  {M.}~\bibnamefont {Hafezi}}, \bibinfo {author} {\bibfnamefont
  {L.}~\bibnamefont {Lu}}, \bibinfo {author} {\bibfnamefont {M.~C.}\
  \bibnamefont {Rechtsman}}, \bibinfo {author} {\bibfnamefont {D.}~\bibnamefont
  {Schuster}}, \bibinfo {author} {\bibfnamefont {J.}~\bibnamefont {Simon}},
  \bibinfo {author} {\bibfnamefont {O.}~\bibnamefont {Zilberberg}}, \emph
  {et~al.},\ }\href {https://doi.org/10.1103/RevModPhys.91.015006} {\bibfield
  {journal} {\bibinfo  {journal} {Reviews of Modern Physics}\ }\textbf
  {\bibinfo {volume} {91}},\ \bibinfo {pages} {015006} (\bibinfo {year}
  {2019})}\BibitemShut {NoStop}%
\bibitem [{\citenamefont {Li}\ \emph {et~al.}(2023)\citenamefont {Li},
  \citenamefont {Luo},\ and\ \citenamefont
  {Gu}}]{liTopologicalOnchipLasers2023}%
  \BibitemOpen
  \bibfield  {author} {\bibinfo {author} {\bibfnamefont {Z.}~\bibnamefont
  {Li}}, \bibinfo {author} {\bibfnamefont {X.-W.}\ \bibnamefont {Luo}},\ and\
  \bibinfo {author} {\bibfnamefont {Q.}~\bibnamefont {Gu}},\ }\href
  {https://doi.org/10.1063/5.0150421} {\bibfield  {journal} {\bibinfo
  {journal} {APL Photonics}\ }\textbf {\bibinfo {volume} {8}},\ \bibinfo
  {pages} {070901} (\bibinfo {year} {2023})}\BibitemShut {NoStop}%
\bibitem [{\citenamefont {Zeng}\ \emph {et~al.}(2020)\citenamefont {Zeng},
  \citenamefont {Chattopadhyay}, \citenamefont {Zhu}, \citenamefont {Qiang},
  \citenamefont {Li}, \citenamefont {Jin}, \citenamefont {Li}, \citenamefont
  {Davies}, \citenamefont {Linfield}, \citenamefont {Zhang}, \citenamefont
  {Chong},\ and\ \citenamefont {Wang}}]{zengElectrically2020}%
  \BibitemOpen
  \bibfield  {author} {\bibinfo {author} {\bibfnamefont {Y.}~\bibnamefont
  {Zeng}}, \bibinfo {author} {\bibfnamefont {U.}~\bibnamefont {Chattopadhyay}},
  \bibinfo {author} {\bibfnamefont {B.}~\bibnamefont {Zhu}}, \bibinfo {author}
  {\bibfnamefont {B.}~\bibnamefont {Qiang}}, \bibinfo {author} {\bibfnamefont
  {J.}~\bibnamefont {Li}}, \bibinfo {author} {\bibfnamefont {Y.}~\bibnamefont
  {Jin}}, \bibinfo {author} {\bibfnamefont {L.}~\bibnamefont {Li}}, \bibinfo
  {author} {\bibfnamefont {A.~G.}\ \bibnamefont {Davies}}, \bibinfo {author}
  {\bibfnamefont {E.~H.}\ \bibnamefont {Linfield}}, \bibinfo {author}
  {\bibfnamefont {B.}~\bibnamefont {Zhang}}, \bibinfo {author} {\bibfnamefont
  {Y.}~\bibnamefont {Chong}},\ and\ \bibinfo {author} {\bibfnamefont {Q.~J.}\
  \bibnamefont {Wang}},\ }\href {https://doi.org/10.1038/s41586-020-1981-x}
  {\bibfield  {journal} {\bibinfo  {journal} {Nature}\ }\textbf {\bibinfo
  {volume} {578}},\ \bibinfo {pages} {246} (\bibinfo {year}
  {2020})}\BibitemShut {NoStop}%
\bibitem [{\citenamefont {Yang}\ \emph {et~al.}(2022)\citenamefont {Yang},
  \citenamefont {Li}, \citenamefont {Gao},\ and\ \citenamefont
  {Lu}}]{yangTopologicalcavity2022}%
  \BibitemOpen
  \bibfield  {author} {\bibinfo {author} {\bibfnamefont {L.}~\bibnamefont
  {Yang}}, \bibinfo {author} {\bibfnamefont {G.}~\bibnamefont {Li}}, \bibinfo
  {author} {\bibfnamefont {X.}~\bibnamefont {Gao}},\ and\ \bibinfo {author}
  {\bibfnamefont {L.}~\bibnamefont {Lu}},\ }\href
  {https://doi.org/10.1038/s41566-022-00972-6} {\bibfield  {journal} {\bibinfo
  {journal} {Nature Photonics}\ }\textbf {\bibinfo {volume} {16}},\ \bibinfo
  {pages} {279} (\bibinfo {year} {2022})}\BibitemShut {NoStop}%
\bibitem [{\citenamefont {Amelio}\ and\ \citenamefont
  {Carusotto}(2020)}]{amelioTheoryCoherenceTopological2020}%
  \BibitemOpen
  \bibfield  {author} {\bibinfo {author} {\bibfnamefont {I.}~\bibnamefont
  {Amelio}}\ and\ \bibinfo {author} {\bibfnamefont {I.}~\bibnamefont
  {Carusotto}},\ }\href
  {https://doi.org/https://doi.org/10.1103/PhysRevX.10.041060} {\bibfield
  {journal} {\bibinfo  {journal} {Physical Review X}\ }\textbf {\bibinfo
  {volume} {10}},\ \bibinfo {pages} {041060} (\bibinfo {year}
  {2020})}\BibitemShut {NoStop}%
\bibitem [{\citenamefont {Deligiannis}\ \emph {et~al.}(2022)\citenamefont
  {Deligiannis}, \citenamefont {Fontaine}, \citenamefont {Squizzato},
  \citenamefont {Richard}, \citenamefont {Ravets}, \citenamefont {Bloch},
  \citenamefont {Minguzzi},\ and\ \citenamefont
  {Canet}}]{PhysRevResearch.4.043207}%
  \BibitemOpen
  \bibfield  {author} {\bibinfo {author} {\bibfnamefont {K.}~\bibnamefont
  {Deligiannis}}, \bibinfo {author} {\bibfnamefont {Q.}~\bibnamefont
  {Fontaine}}, \bibinfo {author} {\bibfnamefont {D.}~\bibnamefont {Squizzato}},
  \bibinfo {author} {\bibfnamefont {M.}~\bibnamefont {Richard}}, \bibinfo
  {author} {\bibfnamefont {S.}~\bibnamefont {Ravets}}, \bibinfo {author}
  {\bibfnamefont {J.}~\bibnamefont {Bloch}}, \bibinfo {author} {\bibfnamefont
  {A.}~\bibnamefont {Minguzzi}},\ and\ \bibinfo {author} {\bibfnamefont
  {L.}~\bibnamefont {Canet}},\ }\href
  {https://doi.org/10.1103/PhysRevResearch.4.043207} {\bibfield  {journal}
  {\bibinfo  {journal} {Phys. Rev. Res.}\ }\textbf {\bibinfo {volume} {4}},\
  \bibinfo {pages} {043207} (\bibinfo {year} {2022})}\BibitemShut {NoStop}%
\bibitem [{\citenamefont {Ahmad}\ \emph {et~al.}(2025)\citenamefont {Ahmad},
  \citenamefont {Dai}, \citenamefont {Feng}, \citenamefont {Chen},
  \citenamefont {Mohmaed}, \citenamefont {Khan}, \citenamefont {Hao},
  \citenamefont {Wang}, \citenamefont {Iqbal},\ and\ \citenamefont
  {Mehvish}}]{Review2025LiquidCrystal}%
  \BibitemOpen
  \bibfield  {author} {\bibinfo {author} {\bibfnamefont {A.}~\bibnamefont
  {Ahmad}}, \bibinfo {author} {\bibfnamefont {H.}~\bibnamefont {Dai}}, \bibinfo
  {author} {\bibfnamefont {S.}~\bibnamefont {Feng}}, \bibinfo {author}
  {\bibfnamefont {Z.}~\bibnamefont {Chen}}, \bibinfo {author} {\bibfnamefont
  {Z.}~\bibnamefont {Mohmaed}}, \bibinfo {author} {\bibfnamefont {A.~A.}\
  \bibnamefont {Khan}}, \bibinfo {author} {\bibfnamefont {X.}~\bibnamefont
  {Hao}}, \bibinfo {author} {\bibfnamefont {Y.}~\bibnamefont {Wang}}, \bibinfo
  {author} {\bibfnamefont {N.}~\bibnamefont {Iqbal}},\ and\ \bibinfo {author}
  {\bibfnamefont {D.}~\bibnamefont {Mehvish}},\ }\href
  {https://doi.org/10.1039/D4TC04871G} {\bibfield  {journal} {\bibinfo
  {journal} {J. Mater. Chem. C}\ }\textbf {\bibinfo {volume} {13}},\ \bibinfo
  {pages} {2606} (\bibinfo {year} {2025})}\BibitemShut {NoStop}%
\bibitem [{\citenamefont {Zhang}\ \emph {et~al.}(2024)\citenamefont {Zhang}
  \emph {et~al.}}]{Zhang2024Perovskite}%
  \BibitemOpen
  \bibfield  {author} {\bibinfo {author} {\bibfnamefont {Y.}~\bibnamefont
  {Zhang}} \emph {et~al.},\ }\href
  {https://doi.org/10.1016/j.jlumin.2024.120946} {\bibfield  {journal}
  {\bibinfo  {journal} {Journal of Luminescence}\ }\textbf {\bibinfo {volume}
  {277}},\ \bibinfo {pages} {120946} (\bibinfo {year} {2024})}\BibitemShut
  {NoStop}%
\bibitem [{\citenamefont {Prokopeva}\ \emph {et~al.}(2024)\citenamefont
  {Prokopeva}, \citenamefont {Fruhling}, \citenamefont {Chowdhury},\ and\
  \citenamefont {Kildishev}}]{Prokopeva2024Modeling}%
  \BibitemOpen
  \bibfield  {author} {\bibinfo {author} {\bibfnamefont {L.~J.}\ \bibnamefont
  {Prokopeva}}, \bibinfo {author} {\bibfnamefont {C.}~\bibnamefont {Fruhling}},
  \bibinfo {author} {\bibfnamefont {S.}~\bibnamefont {Chowdhury}},\ and\
  \bibinfo {author} {\bibfnamefont {A.}~\bibnamefont {Kildishev}},\ }in\ \href
  {https://doi.org/10.1117/12.3028194} {\emph {\bibinfo {booktitle}
  {Proceedings of SPIE 13110, Active Photonic Platforms}}}\ (\bibinfo
  {organization} {SPIE},\ \bibinfo {year} {2024})\ p.\ \bibinfo {pages}
  {1311004}\BibitemShut {NoStop}%
\bibitem [{\citenamefont {Dipold}\ \emph {et~al.}(2024)\citenamefont {Dipold},
  \citenamefont {Kassab},\ and\ \citenamefont {Wetter}}]{Dipold2024Infrared}%
  \BibitemOpen
  \bibfield  {author} {\bibinfo {author} {\bibfnamefont {J.}~\bibnamefont
  {Dipold}}, \bibinfo {author} {\bibfnamefont {L.~R.~P.}\ \bibnamefont
  {Kassab}},\ and\ \bibinfo {author} {\bibfnamefont {N.~U.}\ \bibnamefont
  {Wetter}},\ }\bibfield  {journal} {\bibinfo  {journal} {Photonics}\ }\textbf
  {\bibinfo {volume} {11}},\ \href {https://doi.org/10.3390/photonics11100898}
  {10.3390/photonics11100898} (\bibinfo {year} {2024})\BibitemShut {NoStop}%
\bibitem [{\citenamefont {Segev}\ \emph {et~al.}(2013)\citenamefont {Segev},
  \citenamefont {Silberberg},\ and\ \citenamefont
  {Christodoulides}}]{segevAnderson2013}%
  \BibitemOpen
  \bibfield  {author} {\bibinfo {author} {\bibfnamefont {M.}~\bibnamefont
  {Segev}}, \bibinfo {author} {\bibfnamefont {Y.}~\bibnamefont {Silberberg}},\
  and\ \bibinfo {author} {\bibfnamefont {D.~N.}\ \bibnamefont
  {Christodoulides}},\ }\href {https://doi.org/10.1038/nphoton.2013.30}
  {\bibfield  {journal} {\bibinfo  {journal} {Nature Photonics}\ }\textbf
  {\bibinfo {volume} {7}},\ \bibinfo {pages} {197} (\bibinfo {year}
  {2013})}\BibitemShut {NoStop}%
\bibitem [{\citenamefont {Lee}\ \emph {et~al.}(2019)\citenamefont {Lee},
  \citenamefont {Callard}, \citenamefont {Seassal},\ and\ \citenamefont
  {Jeon}}]{leeTamingRandomLasers2019}%
  \BibitemOpen
  \bibfield  {author} {\bibinfo {author} {\bibfnamefont {M.}~\bibnamefont
  {Lee}}, \bibinfo {author} {\bibfnamefont {S.}~\bibnamefont {Callard}},
  \bibinfo {author} {\bibfnamefont {C.}~\bibnamefont {Seassal}},\ and\ \bibinfo
  {author} {\bibfnamefont {H.}~\bibnamefont {Jeon}},\ }\href
  {https://doi.org/10.1038/s41566-019-0407-5} {\bibfield  {journal} {\bibinfo
  {journal} {Nature Photonics}\ }\textbf {\bibinfo {volume} {13}},\ \bibinfo
  {pages} {445} (\bibinfo {year} {2019})}\BibitemShut {NoStop}%
\bibitem [{\citenamefont {Adl}\ \emph {et~al.}(2024)\citenamefont {Adl},
  \citenamefont {{S{\'a}nchez-D{\'i}az}}, \citenamefont {Vescio}, \citenamefont
  {Cirera}, \citenamefont {Garrido}, \citenamefont {Pacheco}, \citenamefont
  {{\.Z}uraw}, \citenamefont {Przypis}, \citenamefont {{\"O}z}, \citenamefont
  {{Mora-Ser{\'o}}}, \citenamefont {{Mart{\'i}nez-Pastor}},\ and\ \citenamefont
  {Su{\'a}rez}}]{adlTailoringSingleModeRandom2024}%
  \BibitemOpen
  \bibfield  {author} {\bibinfo {author} {\bibfnamefont {H.~P.}\ \bibnamefont
  {Adl}}, \bibinfo {author} {\bibfnamefont {J.}~\bibnamefont
  {{S{\'a}nchez-D{\'i}az}}}, \bibinfo {author} {\bibfnamefont {G.}~\bibnamefont
  {Vescio}}, \bibinfo {author} {\bibfnamefont {A.}~\bibnamefont {Cirera}},
  \bibinfo {author} {\bibfnamefont {B.}~\bibnamefont {Garrido}}, \bibinfo
  {author} {\bibfnamefont {F.~A.~V.}\ \bibnamefont {Pacheco}}, \bibinfo
  {author} {\bibfnamefont {W.}~\bibnamefont {{\.Z}uraw}}, \bibinfo {author}
  {\bibfnamefont {{\L}.}~\bibnamefont {Przypis}}, \bibinfo {author}
  {\bibfnamefont {S.}~\bibnamefont {{\"O}z}}, \bibinfo {author} {\bibfnamefont
  {I.}~\bibnamefont {{Mora-Ser{\'o}}}}, \bibinfo {author} {\bibfnamefont
  {J.~P.}\ \bibnamefont {{Mart{\'i}nez-Pastor}}},\ and\ \bibinfo {author}
  {\bibfnamefont {I.}~\bibnamefont {Su{\'a}rez}},\ }\href
  {https://doi.org/https://doi.org/10.1002/adma.202313252} {\bibfield
  {journal} {\bibinfo  {journal} {Advanced Materials}\ }\textbf {\bibinfo
  {volume} {36}},\ \bibinfo {pages} {2313252} (\bibinfo {year}
  {2024})}\BibitemShut {NoStop}%
\bibitem [{\citenamefont {Shawki}\ \emph {et~al.}(2017)\citenamefont {Shawki},
  \citenamefont {Kotb},\ and\ \citenamefont {Khalil}}]{Shawki17}%
  \BibitemOpen
  \bibfield  {author} {\bibinfo {author} {\bibfnamefont {H.}~\bibnamefont
  {Shawki}}, \bibinfo {author} {\bibfnamefont {H.}~\bibnamefont {Kotb}},\ and\
  \bibinfo {author} {\bibfnamefont {D.}~\bibnamefont {Khalil}},\ }\href
  {https://doi.org/10.1364/OL.42.003247} {\bibfield  {journal} {\bibinfo
  {journal} {Opt. Lett.}\ }\textbf {\bibinfo {volume} {42}},\ \bibinfo {pages}
  {3247} (\bibinfo {year} {2017})}\BibitemShut {NoStop}%
\bibitem [{\citenamefont {Dey}\ \emph {et~al.}(2024)\citenamefont {Dey},
  \citenamefont {Pramanik}, \citenamefont {Biswas}, \citenamefont {Kumbhakar},\
  and\ \citenamefont {Kumbhakar}}]{rlbook}%
  \BibitemOpen
  \bibfield  {author} {\bibinfo {author} {\bibfnamefont {A.}~\bibnamefont
  {Dey}}, \bibinfo {author} {\bibfnamefont {A.}~\bibnamefont {Pramanik}},
  \bibinfo {author} {\bibfnamefont {S.}~\bibnamefont {Biswas}}, \bibinfo
  {author} {\bibfnamefont {P.}~\bibnamefont {Kumbhakar}},\ and\ \bibinfo
  {author} {\bibfnamefont {P.}~\bibnamefont {Kumbhakar}},\ }\bibinfo {title}
  {Developments of random laser: Fundamentals and applications},\ in\ \href
  {https://doi.org/10.1007/978-981-99-7145-9_12} {\emph {\bibinfo {booktitle}
  {Handbook of Materials Science, Volume 1: Optical Materials}}}\ (\bibinfo
  {publisher} {Springer Nature Singapore},\ \bibinfo {address} {Singapore},\
  \bibinfo {year} {2024})\ pp.\ \bibinfo {pages} {341--368}\BibitemShut
  {NoStop}%
\bibitem [{\citenamefont {Padiyakkuth}\ \emph {et~al.}(2022)\citenamefont
  {Padiyakkuth}, \citenamefont {Thomas}, \citenamefont {Antoine},\ and\
  \citenamefont {Kalarikkal}}]{rlreviewnew}%
  \BibitemOpen
  \bibfield  {author} {\bibinfo {author} {\bibfnamefont {N.}~\bibnamefont
  {Padiyakkuth}}, \bibinfo {author} {\bibfnamefont {S.}~\bibnamefont {Thomas}},
  \bibinfo {author} {\bibfnamefont {R.}~\bibnamefont {Antoine}},\ and\ \bibinfo
  {author} {\bibfnamefont {N.}~\bibnamefont {Kalarikkal}},\ }\href
  {https://doi.org/10.1039/D2MA00221C} {\bibfield  {journal} {\bibinfo
  {journal} {Mater. Adv.}\ }\textbf {\bibinfo {volume} {3}},\ \bibinfo {pages}
  {6687} (\bibinfo {year} {2022})}\BibitemShut {NoStop}%
\bibitem [{\citenamefont {Li}\ \emph {et~al.}(2009)\citenamefont {Li},
  \citenamefont {Chu}, \citenamefont {Jain},\ and\ \citenamefont
  {Shen}}]{liTopologicalAndersonInsulator2009}%
  \BibitemOpen
  \bibfield  {author} {\bibinfo {author} {\bibfnamefont {J.}~\bibnamefont
  {Li}}, \bibinfo {author} {\bibfnamefont {R.-L.}\ \bibnamefont {Chu}},
  \bibinfo {author} {\bibfnamefont {J.~K.}\ \bibnamefont {Jain}},\ and\
  \bibinfo {author} {\bibfnamefont {S.-Q.}\ \bibnamefont {Shen}},\ }\href
  {https://doi.org/10.1103/PhysRevLett.102.136806} {\bibfield  {journal}
  {\bibinfo  {journal} {Physical Review Letters}\ }\textbf {\bibinfo {volume}
  {102}},\ \bibinfo {pages} {136806} (\bibinfo {year} {2009})}\BibitemShut
  {NoStop}%
\bibitem [{\citenamefont {Groth}\ \emph {et~al.}(2009)\citenamefont {Groth},
  \citenamefont {Wimmer}, \citenamefont {Akhmerov}, \citenamefont
  {Tworzyd{\l}o},\ and\ \citenamefont
  {Beenakker}}]{grothTheoryTopologicalAnderson2009}%
  \BibitemOpen
  \bibfield  {author} {\bibinfo {author} {\bibfnamefont {C.~W.}\ \bibnamefont
  {Groth}}, \bibinfo {author} {\bibfnamefont {M.}~\bibnamefont {Wimmer}},
  \bibinfo {author} {\bibfnamefont {A.~R.}\ \bibnamefont {Akhmerov}}, \bibinfo
  {author} {\bibfnamefont {J.}~\bibnamefont {Tworzyd{\l}o}},\ and\ \bibinfo
  {author} {\bibfnamefont {C.~W.~J.}\ \bibnamefont {Beenakker}},\ }\bibfield
  {journal} {\bibinfo  {journal} {Physical Review Letters}\ }\href
  {https://doi.org/10.1103/PhysRevLett.103.196805}
  {10.1103/PhysRevLett.103.196805} (\bibinfo {year} {2009})\BibitemShut
  {NoStop}%
\bibitem [{\citenamefont {Shen}(2017)}]{Shen2017}%
  \BibitemOpen
  \bibfield  {author} {\bibinfo {author} {\bibfnamefont {S.-Q.}\ \bibnamefont
  {Shen}},\ }\bibinfo {title} {Topological andersontopological anderson
  insulatorinsulator},\ in\ \href
  {https://doi.org/10.1007/978-981-10-4606-3_12} {\emph {\bibinfo {booktitle}
  {Topological Insulators: Dirac Equation in Condensed Matter}}}\ (\bibinfo
  {publisher} {Springer Singapore},\ \bibinfo {address} {Singapore},\ \bibinfo
  {year} {2017})\ pp.\ \bibinfo {pages} {231--241}\BibitemShut {NoStop}%
\bibitem [{\citenamefont {Xing}\ \emph {et~al.}(2011)\citenamefont {Xing},
  \citenamefont {Zhang},\ and\ \citenamefont
  {Wang}}]{xingTopologicalAndersonInsulator2011}%
  \BibitemOpen
  \bibfield  {author} {\bibinfo {author} {\bibfnamefont {Y.}~\bibnamefont
  {Xing}}, \bibinfo {author} {\bibfnamefont {L.}~\bibnamefont {Zhang}},\ and\
  \bibinfo {author} {\bibfnamefont {J.}~\bibnamefont {Wang}},\ }\href
  {https://doi.org/10.1103/PhysRevB.84.035110} {\bibfield  {journal} {\bibinfo
  {journal} {Physical Review B}\ }\textbf {\bibinfo {volume} {84}},\ \bibinfo
  {pages} {035110} (\bibinfo {year} {2011})}\BibitemShut {NoStop}%
\bibitem [{\citenamefont {Jiang}\ \emph {et~al.}(2009)\citenamefont {Jiang},
  \citenamefont {Wang}, \citenamefont {Sun},\ and\ \citenamefont
  {Xie}}]{jiangNumericalStudyTopological2009}%
  \BibitemOpen
  \bibfield  {author} {\bibinfo {author} {\bibfnamefont {H.}~\bibnamefont
  {Jiang}}, \bibinfo {author} {\bibfnamefont {L.}~\bibnamefont {Wang}},
  \bibinfo {author} {\bibfnamefont {Q.-f.}\ \bibnamefont {Sun}},\ and\ \bibinfo
  {author} {\bibfnamefont {X.~C.}\ \bibnamefont {Xie}},\ }\href
  {https://doi.org/10.1103/PhysRevB.80.165316} {\bibfield  {journal} {\bibinfo
  {journal} {Physical Review B}\ }\textbf {\bibinfo {volume} {80}},\ \bibinfo
  {pages} {165316} (\bibinfo {year} {2009})}\BibitemShut {NoStop}%
\bibitem [{\citenamefont {Bernevig}\ \emph {et~al.}(2006)\citenamefont
  {Bernevig}, \citenamefont {Hughes},\ and\ \citenamefont
  {Zhang}}]{bernevigQuantumSpinHall2006}%
  \BibitemOpen
  \bibfield  {author} {\bibinfo {author} {\bibfnamefont {B.~A.}\ \bibnamefont
  {Bernevig}}, \bibinfo {author} {\bibfnamefont {T.~L.}\ \bibnamefont
  {Hughes}},\ and\ \bibinfo {author} {\bibfnamefont {S.-C.}\ \bibnamefont
  {Zhang}},\ }\href {https://doi.org/10.1126/science.1133734} {\bibfield
  {journal} {\bibinfo  {journal} {Science}\ }\textbf {\bibinfo {volume}
  {314}},\ \bibinfo {pages} {1757} (\bibinfo {year} {2006})}\BibitemShut
  {NoStop}%
\bibitem [{\citenamefont {St{\"u}tzer}\ \emph {et~al.}(2018)\citenamefont
  {St{\"u}tzer}, \citenamefont {Plotnik}, \citenamefont {Lumer}, \citenamefont
  {Titum}, \citenamefont {Lindner}, \citenamefont {Segev}, \citenamefont
  {Rechtsman},\ and\ \citenamefont
  {Szameit}}]{stutzerPhotonicTopologicalAnderson2018}%
  \BibitemOpen
  \bibfield  {author} {\bibinfo {author} {\bibfnamefont {S.}~\bibnamefont
  {St{\"u}tzer}}, \bibinfo {author} {\bibfnamefont {Y.}~\bibnamefont
  {Plotnik}}, \bibinfo {author} {\bibfnamefont {Y.}~\bibnamefont {Lumer}},
  \bibinfo {author} {\bibfnamefont {P.}~\bibnamefont {Titum}}, \bibinfo
  {author} {\bibfnamefont {N.~H.}\ \bibnamefont {Lindner}}, \bibinfo {author}
  {\bibfnamefont {M.}~\bibnamefont {Segev}}, \bibinfo {author} {\bibfnamefont
  {M.~C.}\ \bibnamefont {Rechtsman}},\ and\ \bibinfo {author} {\bibfnamefont
  {A.}~\bibnamefont {Szameit}},\ }\href
  {https://doi.org/10.1038/s41586-018-0536-0} {\bibfield  {journal} {\bibinfo
  {journal} {Nature}\ }\textbf {\bibinfo {volume} {560}},\ \bibinfo {pages}
  {461} (\bibinfo {year} {2018})}\BibitemShut {NoStop}%
\bibitem [{\citenamefont {Liu}\ \emph {et~al.}(2020)\citenamefont {Liu},
  \citenamefont {Yang}, \citenamefont {Ren}, \citenamefont {Xue}, \citenamefont
  {Lin}, \citenamefont {Hu}, \citenamefont {Sun}, \citenamefont {Peng},
  \citenamefont {Zhou}, \citenamefont {Chong},\ and\ \citenamefont
  {Zhang}}]{liuTopologicalAndersonInsulator2020}%
  \BibitemOpen
  \bibfield  {author} {\bibinfo {author} {\bibfnamefont {G.-G.}\ \bibnamefont
  {Liu}}, \bibinfo {author} {\bibfnamefont {Y.}~\bibnamefont {Yang}}, \bibinfo
  {author} {\bibfnamefont {X.}~\bibnamefont {Ren}}, \bibinfo {author}
  {\bibfnamefont {H.}~\bibnamefont {Xue}}, \bibinfo {author} {\bibfnamefont
  {X.}~\bibnamefont {Lin}}, \bibinfo {author} {\bibfnamefont {Y.-H.}\
  \bibnamefont {Hu}}, \bibinfo {author} {\bibfnamefont {H.-x.}\ \bibnamefont
  {Sun}}, \bibinfo {author} {\bibfnamefont {B.}~\bibnamefont {Peng}}, \bibinfo
  {author} {\bibfnamefont {P.}~\bibnamefont {Zhou}}, \bibinfo {author}
  {\bibfnamefont {Y.}~\bibnamefont {Chong}},\ and\ \bibinfo {author}
  {\bibfnamefont {B.}~\bibnamefont {Zhang}},\ }\href
  {https://doi.org/10.1103/PhysRevLett.125.133603} {\bibfield  {journal}
  {\bibinfo  {journal} {Physical Review Letters}\ }\textbf {\bibinfo {volume}
  {125}},\ \bibinfo {pages} {133603} (\bibinfo {year} {2020})}\BibitemShut
  {NoStop}%
\bibitem [{\citenamefont {Chen}\ \emph {et~al.}(2024)\citenamefont {Chen},
  \citenamefont {Gao}, \citenamefont {Cui}, \citenamefont {Mo}, \citenamefont
  {Chen}, \citenamefont {Zhang}, \citenamefont {Chan},\ and\ \citenamefont
  {Dong}}]{chenRealizationTimeReversalInvariant2024}%
  \BibitemOpen
  \bibfield  {author} {\bibinfo {author} {\bibfnamefont {X.-D.}\ \bibnamefont
  {Chen}}, \bibinfo {author} {\bibfnamefont {Z.-X.}\ \bibnamefont {Gao}},
  \bibinfo {author} {\bibfnamefont {X.}~\bibnamefont {Cui}}, \bibinfo {author}
  {\bibfnamefont {H.-C.}\ \bibnamefont {Mo}}, \bibinfo {author} {\bibfnamefont
  {W.-J.}\ \bibnamefont {Chen}}, \bibinfo {author} {\bibfnamefont {R.-Y.}\
  \bibnamefont {Zhang}}, \bibinfo {author} {\bibfnamefont {C.~T.}\ \bibnamefont
  {Chan}},\ and\ \bibinfo {author} {\bibfnamefont {J.-W.}\ \bibnamefont
  {Dong}},\ }\href {https://doi.org/10.1103/PhysRevLett.133.133802} {\bibfield
  {journal} {\bibinfo  {journal} {Physical Review Letters}\ }\textbf {\bibinfo
  {volume} {133}},\ \bibinfo {pages} {133802} (\bibinfo {year}
  {2024})}\BibitemShut {NoStop}%
\bibitem [{\citenamefont {Assun{\c c}{\~a}o}\ \emph {et~al.}(2024)\citenamefont
  {Assun{\c c}{\~a}o}, \citenamefont {Ferreira},\ and\ \citenamefont
  {Lewenkopf}}]{assuncaoPhaseTransitionsScale2024}%
  \BibitemOpen
  \bibfield  {author} {\bibinfo {author} {\bibfnamefont {B.~D.}\ \bibnamefont
  {Assun{\c c}{\~a}o}}, \bibinfo {author} {\bibfnamefont {G.~J.}\ \bibnamefont
  {Ferreira}},\ and\ \bibinfo {author} {\bibfnamefont {C.~H.}\ \bibnamefont
  {Lewenkopf}},\ }\href {https://doi.org/10.1103/PhysRevB.109.L201102}
  {\bibfield  {journal} {\bibinfo  {journal} {Physical Review B}\ }\textbf
  {\bibinfo {volume} {109}},\ \bibinfo {pages} {L201102} (\bibinfo {year}
  {2024})}\BibitemShut {NoStop}%
\bibitem [{\citenamefont {Wu}\ \emph {et~al.}(2022)\citenamefont {Wu},
  \citenamefont {Tang}, \citenamefont {Zhang},\ and\ \citenamefont
  {Zhang}}]{PhysRevA.106.L051301}%
  \BibitemOpen
  \bibfield  {author} {\bibinfo {author} {\bibfnamefont {Y.-P.}\ \bibnamefont
  {Wu}}, \bibinfo {author} {\bibfnamefont {L.-Z.}\ \bibnamefont {Tang}},
  \bibinfo {author} {\bibfnamefont {G.-Q.}\ \bibnamefont {Zhang}},\ and\
  \bibinfo {author} {\bibfnamefont {D.-W.}\ \bibnamefont {Zhang}},\ }\href
  {https://doi.org/10.1103/PhysRevA.106.L051301} {\bibfield  {journal}
  {\bibinfo  {journal} {Phys. Rev. A}\ }\textbf {\bibinfo {volume} {106}},\
  \bibinfo {pages} {L051301} (\bibinfo {year} {2022})}\BibitemShut {NoStop}%
\bibitem [{\citenamefont {Qi}\ \emph {et~al.}(2006)\citenamefont {Qi},
  \citenamefont {Wu},\ and\ \citenamefont {Zhang}}]{Qi2006}%
  \BibitemOpen
  \bibfield  {author} {\bibinfo {author} {\bibfnamefont {X.-L.}\ \bibnamefont
  {Qi}}, \bibinfo {author} {\bibfnamefont {Y.-S.}\ \bibnamefont {Wu}},\ and\
  \bibinfo {author} {\bibfnamefont {S.-C.}\ \bibnamefont {Zhang}},\ }\href
  {https://doi.org/10.1103/PhysRevB.74.085308} {\bibfield  {journal} {\bibinfo
  {journal} {Physical Review B}\ }\textbf {\bibinfo {volume} {74}},\ \bibinfo
  {pages} {085308} (\bibinfo {year} {2006})}\BibitemShut {NoStop}%
\bibitem [{\citenamefont {Asb{\'o}th}\ \emph {et~al.}(2016)\citenamefont
  {Asb{\'o}th}, \citenamefont {Oroszl{\'a}ny},\ and\ \citenamefont
  {P{\'a}lyi}}]{Asbóth2016}%
  \BibitemOpen
  \bibfield  {author} {\bibinfo {author} {\bibfnamefont {J.~K.}\ \bibnamefont
  {Asb{\'o}th}}, \bibinfo {author} {\bibfnamefont {L.}~\bibnamefont
  {Oroszl{\'a}ny}},\ and\ \bibinfo {author} {\bibfnamefont {A.}~\bibnamefont
  {P{\'a}lyi}},\ }\bibinfo {title} {Two-dimensional chern insulators: The
  qi-wu-zhang model},\ in\ \href {https://doi.org/10.1007/978-3-319-25607-8_6}
  {\emph {\bibinfo {booktitle} {A Short Course on Topological Insulators: Band
  Structure\- and Edge States in One and Two Dimensions}}}\ (\bibinfo
  {publisher} {Springer International Publishing},\ \bibinfo {address} {Cham},\
  \bibinfo {year} {2016})\ pp.\ \bibinfo {pages} {85--98}\BibitemShut {NoStop}%
\bibitem [{\citenamefont {Evers}\ and\ \citenamefont {Mirlin}(2008)}]{IPR1}%
  \BibitemOpen
  \bibfield  {author} {\bibinfo {author} {\bibfnamefont {F.}~\bibnamefont
  {Evers}}\ and\ \bibinfo {author} {\bibfnamefont {A.~D.}\ \bibnamefont
  {Mirlin}},\ }\href {https://doi.org/10.1103/RevModPhys.80.1355} {\bibfield
  {journal} {\bibinfo  {journal} {Rev. Mod. Phys.}\ }\textbf {\bibinfo {volume}
  {80}},\ \bibinfo {pages} {1355} (\bibinfo {year} {2008})}\BibitemShut
  {NoStop}%
\bibitem [{\citenamefont {Hikami}(1986)}]{IPR2}%
  \BibitemOpen
  \bibfield  {author} {\bibinfo {author} {\bibfnamefont {S.}~\bibnamefont
  {Hikami}},\ }\href {https://doi.org/10.1143/PTP.76.1210} {\bibfield
  {journal} {\bibinfo  {journal} {Progress of Theoretical Physics}\ }\textbf
  {\bibinfo {volume} {76}},\ \bibinfo {pages} {1210} (\bibinfo {year}
  {1986})},\ \Eprint
  {https://arxiv.org/abs/https://academic.oup.com/ptp/article-pdf/76/6/1210/5344609/76-6-1210.pdf}
  {https://academic.oup.com/ptp/article-pdf/76/6/1210/5344609/76-6-1210.pdf}
  \BibitemShut {NoStop}%
\bibitem [{\citenamefont {Secl\`{i}}\ and\ \citenamefont
  {Carusotto}(2019)}]{Secli:19}%
  \BibitemOpen
  \bibfield  {author} {\bibinfo {author} {\bibfnamefont {M.}~\bibnamefont
  {Secl\`{i}}}\ and\ \bibinfo {author} {\bibfnamefont {I.}~\bibnamefont
  {Carusotto}},\ }in\ \href
  {https://opg.optica.org/abstract.cfm?URI=EQEC-2019-pd_2_5} {\emph {\bibinfo
  {booktitle} {2019 Conference on Lasers and Electro-Optics Europe and European
  Quantum Electronics Conference}}}\ (\bibinfo  {publisher} {Optica Publishing
  Group},\ \bibinfo {year} {2019})\ p.\ \bibinfo {pages} {pd\_2\_5}\BibitemShut
  {NoStop}%
\bibitem [{\citenamefont {Secl{\`i}}\ \emph {et~al.}(2019)\citenamefont
  {Secl{\`i}}, \citenamefont {Capone},\ and\ \citenamefont
  {Carusotto}}]{secliTheoryChiralEdge2019}%
  \BibitemOpen
  \bibfield  {author} {\bibinfo {author} {\bibfnamefont {M.}~\bibnamefont
  {Secl{\`i}}}, \bibinfo {author} {\bibfnamefont {M.}~\bibnamefont {Capone}},\
  and\ \bibinfo {author} {\bibfnamefont {I.}~\bibnamefont {Carusotto}},\ }\href
  {https://doi.org/10.1103/PhysRevResearch.1.033148} {\bibfield  {journal}
  {\bibinfo  {journal} {Physical Review Research}\ }\textbf {\bibinfo {volume}
  {1}},\ \bibinfo {pages} {033148} (\bibinfo {year} {2019})}\BibitemShut
  {NoStop}%
\bibitem [{\citenamefont {Born}\ and\ \citenamefont
  {Wolf}(2019)}]{Born_Wolf_2019}%
  \BibitemOpen
  \bibfield  {author} {\bibinfo {author} {\bibfnamefont {M.}~\bibnamefont
  {Born}}\ and\ \bibinfo {author} {\bibfnamefont {E.}~\bibnamefont {Wolf}},\
  }\href@noop {} {\emph {\bibinfo {title} {Principles of Optics: 60th
  Anniversary Edition}}},\ \bibinfo {edition} {7th}\ ed.\ (\bibinfo
  {publisher} {Cambridge University Press},\ \bibinfo {year}
  {2019})\BibitemShut {NoStop}%
\bibitem [{\citenamefont {Deng}\ \emph {et~al.}(2007)\citenamefont {Deng},
  \citenamefont {Solomon}, \citenamefont {Hey}, \citenamefont {Ploog},\ and\
  \citenamefont {Yamamoto}}]{PhysRevLett.99.126403}%
  \BibitemOpen
  \bibfield  {author} {\bibinfo {author} {\bibfnamefont {H.}~\bibnamefont
  {Deng}}, \bibinfo {author} {\bibfnamefont {G.~S.}\ \bibnamefont {Solomon}},
  \bibinfo {author} {\bibfnamefont {R.}~\bibnamefont {Hey}}, \bibinfo {author}
  {\bibfnamefont {K.~H.}\ \bibnamefont {Ploog}},\ and\ \bibinfo {author}
  {\bibfnamefont {Y.}~\bibnamefont {Yamamoto}},\ }\href
  {https://doi.org/10.1103/PhysRevLett.99.126403} {\bibfield  {journal}
  {\bibinfo  {journal} {Phys. Rev. Lett.}\ }\textbf {\bibinfo {volume} {99}},\
  \bibinfo {pages} {126403} (\bibinfo {year} {2007})}\BibitemShut {NoStop}%
\bibitem [{\citenamefont {Fontaine}\ \emph {et~al.}(2022)\citenamefont
  {Fontaine}, \citenamefont {Squizzato}, \citenamefont {Baboux}, \citenamefont
  {Amelio}, \citenamefont {Lema{\^i}tre}, \citenamefont {Morassi},
  \citenamefont {Sagnes}, \citenamefont {Le~Gratiet}, \citenamefont {Harouri},
  \citenamefont {Wouters}, \citenamefont {Carusotto}, \citenamefont {Amo},
  \citenamefont {Richard}, \citenamefont {Minguzzi}, \citenamefont {Canet},
  \citenamefont {Ravets},\ and\ \citenamefont {Bloch}}]{fontaineKardar2022}%
  \BibitemOpen
  \bibfield  {author} {\bibinfo {author} {\bibfnamefont {Q.}~\bibnamefont
  {Fontaine}}, \bibinfo {author} {\bibfnamefont {D.}~\bibnamefont {Squizzato}},
  \bibinfo {author} {\bibfnamefont {F.}~\bibnamefont {Baboux}}, \bibinfo
  {author} {\bibfnamefont {I.}~\bibnamefont {Amelio}}, \bibinfo {author}
  {\bibfnamefont {A.}~\bibnamefont {Lema{\^i}tre}}, \bibinfo {author}
  {\bibfnamefont {M.}~\bibnamefont {Morassi}}, \bibinfo {author} {\bibfnamefont
  {I.}~\bibnamefont {Sagnes}}, \bibinfo {author} {\bibfnamefont
  {L.}~\bibnamefont {Le~Gratiet}}, \bibinfo {author} {\bibfnamefont
  {A.}~\bibnamefont {Harouri}}, \bibinfo {author} {\bibfnamefont
  {M.}~\bibnamefont {Wouters}}, \bibinfo {author} {\bibfnamefont
  {I.}~\bibnamefont {Carusotto}}, \bibinfo {author} {\bibfnamefont
  {A.}~\bibnamefont {Amo}}, \bibinfo {author} {\bibfnamefont {M.}~\bibnamefont
  {Richard}}, \bibinfo {author} {\bibfnamefont {A.}~\bibnamefont {Minguzzi}},
  \bibinfo {author} {\bibfnamefont {L.}~\bibnamefont {Canet}}, \bibinfo
  {author} {\bibfnamefont {S.}~\bibnamefont {Ravets}},\ and\ \bibinfo {author}
  {\bibfnamefont {J.}~\bibnamefont {Bloch}},\ }\href
  {https://doi.org/10.1038/s41586-022-05001-8} {\bibfield  {journal} {\bibinfo
  {journal} {Nature}\ }\textbf {\bibinfo {volume} {608}},\ \bibinfo {pages}
  {687} (\bibinfo {year} {2022})}\BibitemShut {NoStop}%
\bibitem [{\citenamefont {Wu}\ \emph {et~al.}(2026)\citenamefont {Wu},
  \citenamefont {Ding}, \citenamefont {Yan}, \citenamefont {Jia}, \citenamefont
  {Yu}, \citenamefont {Ni},\ and\ \citenamefont {Chen}}]{solidstate}%
  \BibitemOpen
  \bibfield  {author} {\bibinfo {author} {\bibfnamefont {B.}~\bibnamefont
  {Wu}}, \bibinfo {author} {\bibfnamefont {J.}~\bibnamefont {Ding}}, \bibinfo
  {author} {\bibfnamefont {W.}~\bibnamefont {Yan}}, \bibinfo {author}
  {\bibfnamefont {Y.}~\bibnamefont {Jia}}, \bibinfo {author} {\bibfnamefont
  {H.}~\bibnamefont {Yu}}, \bibinfo {author} {\bibfnamefont {X.}~\bibnamefont
  {Ni}},\ and\ \bibinfo {author} {\bibfnamefont {F.}~\bibnamefont {Chen}},\
  }\href {https://doi.org/10.1038/s41467-025-68173-7} {\bibfield  {journal}
  {\bibinfo  {journal} {Nature Communications}\ }\textbf {\bibinfo {volume}
  {17}},\ \bibinfo {pages} {1426} (\bibinfo {year} {2026})}\BibitemShut
  {NoStop}%
\end{thebibliography}

\begin{thebibliography}{9}%
\makeatletter
\providecommand \@ifxundefined [1]{%
 \@ifx{#1\undefined}
}%
\providecommand \@ifnum [1]{%
 \ifnum #1\expandafter \@firstoftwo
 \else \expandafter \@secondoftwo
 \fi
}%
\providecommand \@ifx [1]{%
 \ifx #1\expandafter \@firstoftwo
 \else \expandafter \@secondoftwo
 \fi
}%
\providecommand \natexlab [1]{#1}%
\providecommand \enquote  [1]{``#1''}%
\providecommand \bibnamefont  [1]{#1}%
\providecommand \bibfnamefont [1]{#1}%
\providecommand \citenamefont [1]{#1}%
\providecommand \href@noop [0]{\@secondoftwo}%
\providecommand \href [0]{\begingroup \@sanitize@url \@href}%
\providecommand \@href[1]{\@@startlink{#1}\@@href}%
\providecommand \@@href[1]{\endgroup#1\@@endlink}%
\providecommand \@sanitize@url [0]{\catcode `\\12\catcode `\$12\catcode
  `\&12\catcode `\#12\catcode `\^12\catcode `\_12\catcode `\%12\relax}%
\providecommand \@@startlink[1]{}%
\providecommand \@@endlink[0]{}%
\providecommand \url  [0]{\begingroup\@sanitize@url \@url }%
\providecommand \@url [1]{\endgroup\@href {#1}{\urlprefix }}%
\providecommand \urlprefix  [0]{URL }%
\providecommand \Eprint [0]{\href }%
\providecommand \doibase [0]{https://doi.org/}%
\providecommand \selectlanguage [0]{\@gobble}%
\providecommand \bibinfo  [0]{\@secondoftwo}%
\providecommand \bibfield  [0]{\@secondoftwo}%
\providecommand \translation [1]{[#1]}%
\providecommand \BibitemOpen [0]{}%
\providecommand \bibitemStop [0]{}%
\providecommand \bibitemNoStop [0]{.\EOS\space}%
\providecommand \EOS [0]{\spacefactor3000\relax}%
\providecommand \BibitemShut  [1]{\csname bibitem#1\endcsname}%
\let\auto@bib@innerbib\@empty
%</preamble>
\bibitem [{\citenamefont {Li}\ \emph {et~al.}(2009)\citenamefont {Li},
  \citenamefont {Chu}, \citenamefont {Jain},\ and\ \citenamefont
  {Shen}}]{liTopologicalAndersonInsulator2009S}%
  \BibitemOpen
  \bibfield  {author} {\bibinfo {author} {\bibfnamefont {J.}~\bibnamefont
  {Li}}, \bibinfo {author} {\bibfnamefont {R.-L.}\ \bibnamefont {Chu}},
  \bibinfo {author} {\bibfnamefont {J.~K.}\ \bibnamefont {Jain}},\ and\
  \bibinfo {author} {\bibfnamefont {S.-Q.}\ \bibnamefont {Shen}},\ }\bibfield
  {title} {\bibinfo {title} {Topological anderson insulator},\ }\href
  {https://doi.org/10.1103/PhysRevLett.102.136806} {\bibfield  {journal}
  {\bibinfo  {journal} {Physical Review Letters}\ }\textbf {\bibinfo {volume}
  {102}},\ \bibinfo {pages} {136806} (\bibinfo {year} {2009})}\BibitemShut
  {NoStop}%
\bibitem [{\citenamefont {St{\"u}tzer}\ \emph {et~al.}(2018)\citenamefont
  {St{\"u}tzer}, \citenamefont {Plotnik}, \citenamefont {Lumer}, \citenamefont
  {Titum}, \citenamefont {Lindner}, \citenamefont {Segev}, \citenamefont
  {Rechtsman},\ and\ \citenamefont
  {Szameit}}]{stutzerPhotonicTopologicalAnderson2018S}%
  \BibitemOpen
  \bibfield  {author} {\bibinfo {author} {\bibfnamefont {S.}~\bibnamefont
  {St{\"u}tzer}}, \bibinfo {author} {\bibfnamefont {Y.}~\bibnamefont
  {Plotnik}}, \bibinfo {author} {\bibfnamefont {Y.}~\bibnamefont {Lumer}},
  \bibinfo {author} {\bibfnamefont {P.}~\bibnamefont {Titum}}, \bibinfo
  {author} {\bibfnamefont {N.~H.}\ \bibnamefont {Lindner}}, \bibinfo {author}
  {\bibfnamefont {M.}~\bibnamefont {Segev}}, \bibinfo {author} {\bibfnamefont
  {M.~C.}\ \bibnamefont {Rechtsman}},\ and\ \bibinfo {author} {\bibfnamefont
  {A.}~\bibnamefont {Szameit}},\ }\bibfield  {title} {\bibinfo {title}
  {Photonic topological anderson insulators},\ }\href
  {https://doi.org/10.1038/s41586-018-0536-0} {\bibfield  {journal} {\bibinfo
  {journal} {Nature}\ }\textbf {\bibinfo {volume} {560}},\ \bibinfo {pages}
  {461} (\bibinfo {year} {2018})}\BibitemShut {NoStop}%
\bibitem [{\citenamefont {Liu}\ \emph {et~al.}(2020)\citenamefont {Liu},
  \citenamefont {Yang}, \citenamefont {Ren}, \citenamefont {Xue}, \citenamefont
  {Lin}, \citenamefont {Hu}, \citenamefont {Sun}, \citenamefont {Peng},
  \citenamefont {Zhou}, \citenamefont {Chong},\ and\ \citenamefont
  {Zhang}}]{liuTopologicalAndersonInsulator2020S}%
  \BibitemOpen
  \bibfield  {author} {\bibinfo {author} {\bibfnamefont {G.-G.}\ \bibnamefont
  {Liu}}, \bibinfo {author} {\bibfnamefont {Y.}~\bibnamefont {Yang}}, \bibinfo
  {author} {\bibfnamefont {X.}~\bibnamefont {Ren}}, \bibinfo {author}
  {\bibfnamefont {H.}~\bibnamefont {Xue}}, \bibinfo {author} {\bibfnamefont
  {X.}~\bibnamefont {Lin}}, \bibinfo {author} {\bibfnamefont {Y.-H.}\
  \bibnamefont {Hu}}, \bibinfo {author} {\bibfnamefont {H.-x.}\ \bibnamefont
  {Sun}}, \bibinfo {author} {\bibfnamefont {B.}~\bibnamefont {Peng}}, \bibinfo
  {author} {\bibfnamefont {P.}~\bibnamefont {Zhou}}, \bibinfo {author}
  {\bibfnamefont {Y.}~\bibnamefont {Chong}},\ and\ \bibinfo {author}
  {\bibfnamefont {B.}~\bibnamefont {Zhang}},\ }\bibfield  {title} {\bibinfo
  {title} {Topological anderson insulator in disordered photonic crystals},\
  }\href {https://doi.org/10.1103/PhysRevLett.125.133603} {\bibfield  {journal}
  {\bibinfo  {journal} {Physical Review Letters}\ }\textbf {\bibinfo {volume}
  {125}},\ \bibinfo {pages} {133603} (\bibinfo {year} {2020})}\BibitemShut
  {NoStop}%
\bibitem [{\citenamefont {Chen}\ \emph {et~al.}(2024)\citenamefont {Chen},
  \citenamefont {Gao}, \citenamefont {Cui}, \citenamefont {Mo}, \citenamefont
  {Chen}, \citenamefont {Zhang}, \citenamefont {Chan},\ and\ \citenamefont
  {Dong}}]{chenRealizationTimeReversalInvariant2024S}%
  \BibitemOpen
  \bibfield  {author} {\bibinfo {author} {\bibfnamefont {X.-D.}\ \bibnamefont
  {Chen}}, \bibinfo {author} {\bibfnamefont {Z.-X.}\ \bibnamefont {Gao}},
  \bibinfo {author} {\bibfnamefont {X.}~\bibnamefont {Cui}}, \bibinfo {author}
  {\bibfnamefont {H.-C.}\ \bibnamefont {Mo}}, \bibinfo {author} {\bibfnamefont
  {W.-J.}\ \bibnamefont {Chen}}, \bibinfo {author} {\bibfnamefont {R.-Y.}\
  \bibnamefont {Zhang}}, \bibinfo {author} {\bibfnamefont {C.~T.}\ \bibnamefont
  {Chan}},\ and\ \bibinfo {author} {\bibfnamefont {J.-W.}\ \bibnamefont
  {Dong}},\ }\bibfield  {title} {\bibinfo {title} {Realization of time-reversal
  invariant photonic topological anderson insulators},\ }\href
  {https://doi.org/10.1103/PhysRevLett.133.133802} {\bibfield  {journal}
  {\bibinfo  {journal} {Physical Review Letters}\ }\textbf {\bibinfo {volume}
  {133}},\ \bibinfo {pages} {133802} (\bibinfo {year} {2024})}\BibitemShut
  {NoStop}%
\bibitem [{\citenamefont {Assun{\c c}{\~a}o}\ \emph {et~al.}(2024)\citenamefont
  {Assun{\c c}{\~a}o}, \citenamefont {Ferreira},\ and\ \citenamefont
  {Lewenkopf}}]{assuncaoPhaseTransitionsScale2024S}%
  \BibitemOpen
  \bibfield  {author} {\bibinfo {author} {\bibfnamefont {B.~D.}\ \bibnamefont
  {Assun{\c c}{\~a}o}}, \bibinfo {author} {\bibfnamefont {G.~J.}\ \bibnamefont
  {Ferreira}},\ and\ \bibinfo {author} {\bibfnamefont {C.~H.}\ \bibnamefont
  {Lewenkopf}},\ }\bibfield  {title} {\bibinfo {title} {Phase transitions and
  scale invariance in topological anderson insulators},\ }\href
  {https://doi.org/10.1103/PhysRevB.109.L201102} {\bibfield  {journal}
  {\bibinfo  {journal} {Physical Review B}\ }\textbf {\bibinfo {volume}
  {109}},\ \bibinfo {pages} {L201102} (\bibinfo {year} {2024})}\BibitemShut
  {NoStop}%
\bibitem [{\citenamefont {Wu}\ \emph {et~al.}(2022)\citenamefont {Wu},
  \citenamefont {Tang}, \citenamefont {Zhang},\ and\ \citenamefont
  {Zhang}}]{PhysRevA.106.L051301S}%
  \BibitemOpen
  \bibfield  {author} {\bibinfo {author} {\bibfnamefont {Y.-P.}\ \bibnamefont
  {Wu}}, \bibinfo {author} {\bibfnamefont {L.-Z.}\ \bibnamefont {Tang}},
  \bibinfo {author} {\bibfnamefont {G.-Q.}\ \bibnamefont {Zhang}},\ and\
  \bibinfo {author} {\bibfnamefont {D.-W.}\ \bibnamefont {Zhang}},\ }\bibfield
  {title} {\bibinfo {title} {Quantized topological anderson-thouless pump},\
  }\href {https://doi.org/10.1103/PhysRevA.106.L051301} {\bibfield  {journal}
  {\bibinfo  {journal} {Phys. Rev. A}\ }\textbf {\bibinfo {volume} {106}},\
  \bibinfo {pages} {L051301} (\bibinfo {year} {2022})}\BibitemShut {NoStop}%
\bibitem [{\citenamefont {Sakurai}\ and\ \citenamefont
  {Napolitano}(2020)}]{Sakurai:2011zzS}%
  \BibitemOpen
  \bibfield  {author} {\bibinfo {author} {\bibfnamefont {J.~J.}\ \bibnamefont
  {Sakurai}}\ and\ \bibinfo {author} {\bibfnamefont {J.}~\bibnamefont
  {Napolitano}},\ }\href {https://doi.org/10.1017/9781108587280} {\emph
  {\bibinfo {title} {{Modern Quantum Mechanics}}}},\ \bibinfo {edition} {3rd}\
  ed.,\ Quantum physics, quantum information and quantum computation\ (\bibinfo
   {publisher} {Cambridge University Press},\ \bibinfo {year}
  {2020})\BibitemShut {NoStop}%
\bibitem [{\citenamefont {Amelio}\ and\ \citenamefont
  {Carusotto}(2020)}]{amelioTheoryCoherenceTopological2020S}%
  \BibitemOpen
  \bibfield  {author} {\bibinfo {author} {\bibfnamefont {I.}~\bibnamefont
  {Amelio}}\ and\ \bibinfo {author} {\bibfnamefont {I.}~\bibnamefont
  {Carusotto}},\ }\bibfield  {title} {\bibinfo {title} {Theory of the coherence
  of topological lasers},\ }\href
  {https://doi.org/https://doi.org/10.1103/PhysRevX.10.041060} {\bibfield
  {journal} {\bibinfo  {journal} {Physical Review X}\ }\textbf {\bibinfo
  {volume} {10}},\ \bibinfo {pages} {041060} (\bibinfo {year}
  {2020})}\BibitemShut {NoStop}%
\bibitem [{\citenamefont {Cui}\ \emph {et~al.}(2022)\citenamefont {Cui},
  \citenamefont {Zhang}, \citenamefont {Zhang},\ and\ \citenamefont
  {Chan}}]{cuiPhotonic2Topological2022S}%
  \BibitemOpen
  \bibfield  {author} {\bibinfo {author} {\bibfnamefont {X.}~\bibnamefont
  {Cui}}, \bibinfo {author} {\bibfnamefont {R.-Y.}\ \bibnamefont {Zhang}},
  \bibinfo {author} {\bibfnamefont {Z.-Q.}\ \bibnamefont {Zhang}},\ and\
  \bibinfo {author} {\bibfnamefont {C.~T.}\ \bibnamefont {Chan}},\ }\bibfield
  {title} {\bibinfo {title} {Photonic z 2 topological anderson insulators},\
  }\href {https://doi.org/10.1103/PhysRevLett.129.043902} {\bibfield  {journal}
  {\bibinfo  {journal} {Physical Review Letters}\ }\textbf {\bibinfo {volume}
  {129}},\ \bibinfo {pages} {043902} (\bibinfo {year} {2022})}\BibitemShut
  {NoStop}%
\end{thebibliography}

%apsrev4-2.bst 2019-01-14 (MD) hand-edited version of apsrev4-1.bst
%Control: key (0)
%Control: author (8) initials jnrlst
%Control: editor formatted (1) identically to author
%Control: production of article title (0) allowed
%Control: page (0) single
%Control: year (1) truncated
%Control: production of eprint (0) enabled
%

\end{document}